\documentclass[a4paper,11pt]{article}
\pdfoutput=1 
\usepackage{jheppub} 
\usepackage{bm,graphicx,color,hyperref,slashed}
\usepackage{amssymb,amsmath}
\usepackage[caption=false]{subfig}
\usepackage[T1]{fontenc}

\graphicspath{{./Figures/}}

\newcommand{\diff}{\mathrm{d}}
\newcommand{\p}{\partial}

\newcommand{\Diff}{{\mathcal{D}}}

\newcommand{\be}{\begin{equation}}      
\newcommand{\ee}{\end{equation}}      
\newcommand{\bea}{\begin{eqnarray}}      
\newcommand{\eea}{\end{eqnarray}}

\newcommand{\im}{\mathrm{i}}

\title{Multi-flavor massless QED$_2$ at finite densities via Lefschetz thimbles}

\author[a]{Yuya Tanizaki}

\affiliation[a]{RIKEN BNL Research Center, Brookhaven National Laboratory, Upton, NY 11973-5000 USA}

\emailAdd{yuya.tanizaki@riken.jp}

\author[b]{Motoi Tachibana}

\affiliation[b]{Department of Physics, Saga University, Saga 840-8502, Japan}

\emailAdd{motoi@cc.saga-u.ac.jp}

\abstract{
We consider multi-flavor massless $(1+1)$-dimensional QED with chemical potentials at finite spatial length and the zero-temperature limit. Its sign problem is solved
using the mean-field calculation with complex saddle points. 
}

\begin{document}
\maketitle
\section{Introduction}\label{sec:introduction}

Path integral formalism has been used in a wide range to study systems of quantum field theory as well as statistical mechanics. The formalism is useful not only for the perturbative calculations but also for the numerical computations
such as the lattice Monte Carlo simulations. 
However, it is well known that the lattice Monte Carlo simulation does not work when the action of the system becomes complex and this is called the sign problem~\cite{PhysRevB.41.9301, Batrouni:1992fj}. It is a big obstacle when we try to compute the high-density nuclear matter based on quantum chromodynamics (QCD)~\cite{Muroya:2003qs}. 

In this paper, we consider $(1+1)$-dimensional quantum electrodynamics (QED$_2$) with $N_f$ massless fermions, which we call multi-flavor massless QED$_2$. It is also known as Schwinger model, which is exactly solvable and mapped to the theory of free massive photons~ \cite{Schwinger:1962tn, Schwinger:1962tp}. 
This model has the sign problem at finite temperature and finite number densities, but its phase structure is studied analytically by Refs.~\cite{Narayanan:2012du, Narayanan:2012qf, Lohmayer:2013eka}. Therefore, multi-flavor massless QED$_2$ has been used for the test of various approaches to the sign problem; the dual formulation is discussed in Ref.~\cite{Gattringer:2015nea}, and the study with matrix product states is used for two-flavor massless and massive QED$_2$ in Ref.~\cite{Banuls:2016gid}. 
Here, we apply the Lefshetz-thimble method to obtain phase diagram of multi-flavor massless QED$_2$, and study its property about the sign problem. 

The Lefshetz-thimble method was originally developed to study the hyperasymptotic behavior of exponential integrals~\cite{pham1983vanishing, Kaminski1994, Howls2271}, and it was utilized in study of  Cherns-Simons theory by Witten \cite{Witten:2010cx, Witten:2010zr, Harlow:2011ny}. 
Hyperasymptotics is a branch of the resurgence theory, and its application to semiclassical analysis has been recently discussed extensively~\cite{Dunne:2012ae,Basar:2013eka,Cherman:2014ofa,
Cherman:2014xia,Dorigoni:2014hea, David:1992za,Felder:2004uy,Marino:2008ya,
Marino:2012zq,Schiappa:2013opa, Behtash:2015kna, Behtash:2015kva, Gukov:2016njj, Gukov:2016tnp, Fujimori:2016ljw, Kozcaz:2016wvy}. 
Lefschetz thimble is the higher-dimensional generalization of steepest descent paths, and it has been applied to the sign problem~\cite{Cristoforetti:2012su, Cristoforetti:2013wha, Cristoforetti:2014gsa, Aarts:2013fpa, Fujii:2013sra, Mukherjee:2014hsa, Aarts:2014nxa, Tanizaki:2014xba, Cherman:2014sba, Tanizaki:2014tua, Kanazawa:2014qma, Tanizaki:2015pua, DiRenzo:2015foa, Fukushima:2015qza, Tsutsui:2015tua, Tanizaki:2015rda, Fujii:2015bua, Fujii:2015vha, Alexandru:2015xva, Hayata:2015lzj, Alexandru:2015sua, Alexandru:2016gsd, Alexandru:2016ejd}. 
In this method, complex saddle points of the action play an important role, and the saddle-point approximation with complex saddles describes the essence of nonperturbative behaviors in some cases; it is used for the pair creation under strong laser pulse in Refs.~\cite{Dumlu:2010ua, Dumlu:2011rr, Dumlu:2011cc}, and complex saddles also describe the strongly-coupled phase of the Gross--Witten--Wadia model~\cite{Buividovich:2015oju, Alvarez:2016rmo}. 

Using Lefschetz thimbles, we show that mean-field calculation with complex saddles becomes exact in multi-flavor massless QED$_2$ at finite spacial length and the zero-temperature limit. 
We identify all the complex saddle points of QED$_2$ at finite chemical potentials, and draw the phase diagram using that information. 
At that stage, we must know the intersection numbers between dual thimbles and the original integration region, and those quantities are identified by solving the gradient flow equation that defines Lefschetz thimbles. 
This is the first non-trivial example of gauge theories with the sign problem that is completely solved by the Lefschetz-thimble method. 

This paper is organized as follows. In Section \ref{sec:2dimQED}, we review the computation of massless QED$_2$. In Section \ref{sec:formalism}, we explain the mean-field approximation in the presence of the sign problem. In Section \ref{sec:calculation}, we compute the complex saddle points and obtain the phase structures concretely for one-, two-, and three-flavor cases. To discuss the phase structure for three-flavor case, however, we need the intersection numbers.  For that purpose, we numerically solve the gradient flow and visualize its structure in Section \ref{sec:gradient_flow}. We summarize our result and discuss some perspectives in Section \ref{sec:conclusion}.

\section{$(1+1)$-dimensional massless QED}\label{sec:2dimQED}

In this section, we review the computation of the massless QED$_2$. This section does not present a new result, and the main purpose is to set the notation and make the paper self-contained. 

\subsection{Setup of the model}
Let us first consider the path-integral expression of multi-flavor massless QED$_2$ on a two-dimensional torus $T^2=[0,\beta]\times [0,L]$. The theory is defined by 
\be
Z(\beta,L,\mu)=\int \Diff A\, \mathrm{e}^{-S_{\mathrm{Maxwell}}[A]}\int\Diff \overline{\psi}\Diff \psi \exp\left(-\sum_{a=1}^{N_f}\int \diff^2x\, \overline{\psi}_a\left[\slashed{D}_A-\mu_a \gamma^0\right]\psi_a\right). 
\ee
Here, $A=A_{\mu} \diff x^{\mu}$ is the $U(1)$ gauge field, $\psi_a$ and $\overline{\psi_a}$ are two-component spinor fields with the flavor index $a=1,\ldots,N_f$, $\slashed{D}_A=\gamma_{\nu}(\p_{\nu}+\im A_{\nu})$ with Gamma matrices $\gamma_{\nu}$, and $\mu_a$ is the chemical potentials for fermion number densities $n_a=\overline{\psi}_a\gamma_0\psi_a$. 
$S_{\mathrm{Maxwell}}$ is the Maxwell action, 
\be
S_{\mathrm{Maxwell}}[A]={1\over 4e^2}\int \diff^2 x (\p_{\mu}A_{\nu}-\p_{\nu}A_{\mu})^2. 
\ee
The theory is now defined on the torus $T^2=[0,\beta]\times [0,L]$ with the thermal boundary condition, i.e., 
\bea
&&A_{\nu}(x^0+n\beta,x^1+m L)\sim A_{\nu}(x^0,x^1),\\ 
&&\psi(x^0+n\beta,x^1+m L)=(-1)^n\psi(x^0,x^1). 
\eea
The ``$\sim$'' in the first equation means that the difference between both sides must be appropriately Dirac quantized. 

We review the computation of the Dirac operator on the torus $T^2$ according to Ref.~\cite{Sachs:1991en}. 
We decompose the gauge field as 
\be
A=\left({2\pi k\over \beta L}x^1+{2\pi\over \beta}h_0\right)\diff x^0+{2\pi\over L}h_1\diff x^1+*\diff \phi+\diff \lambda. 
\ee
The meaning of each variables is as follows: $\lambda$ is a gauge parameter, so it does not affect the computation of physical quantities, such as the partition function. $\phi$ is a $2\pi$-periodic function on torus without the zero-momentum mode, which represents the local fluctuation of the photon field. Since $F_{01}=-{2\pi k/\beta L}+\Delta \phi$, the Maxwell action becomes
\be
S_{\mathrm{Maxwell}}={1\over 2 e^2}\int_{T^2} \diff^2 x \left(\Delta \phi-{2\pi k\over \beta L}\right)^2. 
\ee
The fields $h_0$ and $h_1$ are sometimes called toron fields: they do not create the electric field, but describe the nontrivial holonomy at $\phi=0$ and $k=0$, since 
\be
\left.\exp \left(\im \int_\mathcal{C} A\right)\right|_{\phi=0,k=0}=\mathrm{e}^{2\pi \im (n_0h_0+n_1h_1)}, 
\ee
where $n_0$ and $n_1$ are the number of the windings of the closed curve $\mathcal{C}$ along $x^0$- and $x^1$-directions of $T^2$, respectively ($0\le h_\mu<1$). 
The integer $k$ refers the Chern number, 
\be
{1\over 2\pi}\int_{T^2} \diff A=-k,
\ee
which creates the constant electric field. 
The index theorem tells us that $k$ represents the index of the Dirac operator, and thus we need to take into account only the $k=0$ sector: otherwise, the Dirac zero modes exist and the fermion determinant becomes zero because the fermions are massless. 

Let us restrict ourselves to the case $k=0$ in the following, and assume that the Dirac zero mode does not exist. For simplicity of the computation, we put $\lambda=0$ by gauge fixing. 
By putting 
\be
\widetilde{A}={2\pi\over \beta}h_0\diff x^0+{2\pi\over L}h_1\diff x^1, 
\ee
we get $A=\widetilde{A}+*\diff \phi$. We define the chirality matrix by $\gamma=-\im \gamma^0\gamma^1$, then 
\be
\slashed{D}_A=\mathrm{e}^{-\gamma \phi}\slashed{D}_{\widetilde{A}}\mathrm{e}^{-\gamma \phi}. 
\ee
Therefore, we obtain that 
\be
\mathrm{det}\left(\slashed{D}_A\right)=\mathrm{det}\left(\slashed{D}_{\widetilde{A}}\right)\exp\left(-{1\over 2\pi}\int \diff^2 x \phi (-\Delta \phi)\right). 
\ee
In order to obtain this result, we perform the chiral rotation $\psi\mapsto \mathrm{e}^{\gamma\phi}\psi$ and $\overline{\psi}\mapsto \overline{\psi}\mathrm{e}^{\gamma\phi}$, and use the anomaly equation. 
It is important to notice that the global fluctuations $\widetilde{A}$ and the local fluctuation $\phi$ decouple completely from one another at the $k=0$ sector, thanks to the absence of the Dirac zero modes. 
Now, the partition function reads
\bea
Z(\beta,L,\mu)&=&\int_0^1 \diff h_0\diff h_1\, \prod_{a=1}^{N_f}\mathrm{det}\left(\slashed{D}_{\widetilde{A}}-\mu_a\gamma^0\right)\nonumber\\
&&\times \int \Diff \phi\; \exp\left(-{1\over 2e^2}\int \diff^2 x \phi\left(-\Delta+N_f{e^2\over \pi^2}\right)(-\Delta)\phi\right). 
\eea
The photon field $\phi$ becomes massive due to the chiral anomaly, and it decouples from the integration of $h_0$ and $h_1$ fields. 
In this paper, we are interested only in the dependence on the chemical potentials, and the $h_0$ and $h_1$ integrations are most important. In the following, we simply denote that 
\be
Z(\beta,L,\mu)=\int_0^1 \diff h_0\diff h_1\, \prod_{a=1}^{N_f}\mathrm{det}\left(\slashed{D}_{\widetilde{A}}-\mu_a\gamma^0\right). 
\label{eq:partition_function_01}
\ee
In the next subsection, we will compute the fermion determinant under the background with nontrivial holonomies. 

\subsection{Fermion determinant with nontrivial holonomies}

One can compute the Dirac determinant in a way to keep the global-gauge invariance manifestly by using the zeta-function regularization (see Ref.~\cite{Sachs:1991en} and also Appendix in Ref.~\cite{Langfeld:2011rh}). 
To understand the formula intuitively, we derive the result for the fermion determinant heuristically by considering the fermion spectrum. After the chiral transformation, the fermion operator becomes 
\be
\gamma_0 \slashed{D}_{\widetilde{A}}=
\left(\p_{\tau}+{2\pi \im\over \beta}h_0\right)\bm{1}_{2} +\im\left(\p_1+{2\pi \im\over L}h_1\right)\gamma, 
\ee
where $\gamma=-\im \gamma_0\gamma_1$ and in the chiral notation $\gamma=\sigma_3$. 
Therefore, the spectrum of this operator can be labeled as 
\be
{2\pi \over L}(n\pm h_1)+{2\pi \im \over \beta}h_0 +{(2m+1) \pi\im \over \beta}. 
\ee
By performing the summation over Matsubara frequencies $m$, we obtain the Fermi-Dirac distribution for each fermionic mode:
\be
1+\exp\left[-{2\pi \beta\over L}(n\pm h_1)-2\pi \im h_0\right]. 
\ee
Contribution of the positive chiral fermions gives
\be
\left(1+\mathrm{e}^{-{2\pi\beta\over L}  (n+h_1-1)-2\pi \im h_0}\right) \left(1+\mathrm{e}^{-{2\pi \beta\over L} (n-h_1)+2\pi \im h_0}\right), 
\ee
and that of the negative chiral fermions gives 
\be
\left(1+\mathrm{e}^{-{2\pi\beta\over L}  (n+h_1-1)+2\pi \im h_0}\right) \left(1+\mathrm{e}^{-{2\pi \beta\over L} (n-h_1)-2\pi \im h_0}\right). 
\ee
Each term is invariant under $h_0\mapsto h_0+1$, but it is not the case for $h_1\mapsto h_1+1$. 
Even after taking the product over all spatial momenta $n=1,2,\ldots$, the invariance under the spatial global gauge transformation is lost by 
\be
\mathrm{e}^{-{2\pi \beta\over L} (-h_1)+2\pi \im h_0}\mathrm{e}^{-{2\pi \beta\over L} (-h_1)-2\pi \im h_0}=\mathrm{e}^{{4\pi \beta\over L} h_1}. 
\ee
To compensate this factor, one should multiply 
\be
\mathrm{e}^{-{2\pi\beta\over L}(h_1^2-h_1)}. 
\ee
We introduce the dimensionless chemical potential $\mu'$ and the dimensionless temperature $\tau$ by 
\be
\mu'={L\mu\over 2\pi},\quad \tau={L\over \beta}. 
\ee
The $\mu'$-dependence of each determinant becomes 
\bea
\det(\slashed{D}_{\widetilde{A}}-\mu\gamma_0)&=& \mathrm{e}^{-{2\pi\over \tau}(h_1^2-h_1)}\prod_{n=1}^{\infty}\Bigl\{\left(1+\mathrm{e}^{-{2\pi\over \tau}  \left(n+h_1-1-\mu'\right)-2\pi \im h_0}\right) \left(1+\mathrm{e}^{-{2\pi \over \tau} \left(n-h_1+\mu'\right)+2\pi \im h_0}\right)\nonumber\\
&&\times \left(1+\mathrm{e}^{-{2\pi\over \tau}  \left(n+h_1-1+\mu'\right)+2\pi \im h_0}\right) \left(1+\mathrm{e}^{-{2\pi \over \tau} \left(n-h_1-\mu'\right)-2\pi \im h_0}\right)\Bigr\}. 
\eea
This matches with the correct formula derived in Ref.~\cite{Sachs:1991en} up to an uninteresting field-independent factor. 
We define the dimensionless mean-field free energy $F$ by 
\be
F(h_0,h_1)=-{\tau \over 2\pi}\sum_{a=1}^{N_f}\ln \mathrm{det} \left(\slashed{D}_{\widetilde{A}}-{2\pi\over L}\mu'_a\gamma_0\right).  
\ee
The explicit form of this one-loop effective potential becomes 
\bea
F&=& N_f\left(h_1-{1\over 2}\right)^2\nonumber\\
&&-{\tau \over 2\pi}\sum_{a=1}^{N_f} \sum_{n=1}^{\infty}\Bigl\{ \ln \left(1+\mathrm{e}^{-{2\pi\over \tau}  \left(n+h_1-1-\mu'_a\right)-2\pi \im h_0}\right) 
+\ln \left(1+\mathrm{e}^{-{2\pi \over \tau} \left(n-h_1+\mu'_a\right)+2\pi \im h_0}\right)\nonumber\\
&&\quad+\ln \left(1+\mathrm{e}^{-{2\pi\over \tau}  \left(n+h_1-1+\mu'_a\right)+2\pi \im h_0}\right) 
+\ln\left(1+\mathrm{e}^{-{2\pi \over \tau} \left(n-h_1-\mu'_a\right)-2\pi \im h_0}\right)\Bigr\}.
\label{eq:free_energy_qed2}
\eea
Using this one-loop effective potential, (\ref{eq:partition_function_01}) becomes 
\be
Z=\int_{0}^{1} \diff h_0 \diff h_1 \exp\left[-{2\pi\over \tau}F(h_0,h_1)\right]. 
\label{eq:partition_function_QED2}
\ee
We consider about this exponential integral in the zero-temperature limit $\tau\to 0$. 

\section{Mean-field approximation with complex saddle points}\label{sec:formalism}
In this section, we explain the mean-field approximation when the microscopic theory suffers from the sign problem~\cite{Tanizaki:2015pua}. 

\subsection{Lefschetz-thimble methods and mean-field approximation}\label{sec:complex_mean_field}

Here, we explain the general formalism to apply the mean-field approximation when the theory suffers from the sign problem. That is, we consider to apply the mean-field approximation to the path integral 
\be
Z=\int \Diff \phi \exp\left(-S[\phi]\right), 
\ee
where the classical action $S[\phi]$ takes complex values. To consider the mean-field approximation, we introduce the order-parameter operator $O[\phi]$, and define the constrained free energy density~\cite{KorthalsAltes:1993ca, Fukuda:1974ey} by 
\be
F(\Phi)=-{1\over V}\ln \left[\int \Diff \phi \exp\left(-S[\phi]\right) \delta(O[\phi]-\Phi)\right]. 
\ee
If $S[\phi]$ is real, then $F[\Phi]$ is also real and one can compute phase diagram by taking the minimum of $F(\Phi)$. However, if there is the sign problem, the free energy $F(\Phi)$ is complex, and its physical meaning becomes unclear~\cite{Dumitru:2005ng, Fukushima:2006uv}. 

The partition function and the constrained free energy is related by 
\be
Z=\int_{\mathcal{M}} \diff \Phi \exp\left(-V F(\Phi)\right), 
\ee
where $\mathcal{M}$ is the target space of order parameters $\Phi$. 
This corresponds to (\ref{eq:partition_function_QED2}) in our model, where $\Phi=(h_0,h_1)$ and $\mathcal{M}=(\mathbb{R}/\mathbb{Z})^2$. 
In the mean-field approximation, we would like to evaluate this integral in the limit $V\to \infty$ using the saddle-point approximation. 
Since $F[\Phi]$ is complex, this integral is an multi-dimensional oscillatory integral, and we need a technique to treat it. Here, we use the knowledge of the hyperasymptotic analysis~\cite{pham1983vanishing, Kaminski1994, Howls2271}, which is recently used for the study of sign problem of the lattice Monte Carlo simulation and known as the Lefschetz-thimble method~\cite{Cristoforetti:2012su, Cristoforetti:2013wha, Cristoforetti:2014gsa, Aarts:2013fpa, Fujii:2013sra, Mukherjee:2014hsa, Aarts:2014nxa, Tanizaki:2014xba, Cherman:2014sba, Tanizaki:2014tua, Kanazawa:2014qma, Tanizaki:2015pua, DiRenzo:2015foa, Fukushima:2015qza, Tsutsui:2015tua, Tanizaki:2015rda, Fujii:2015bua, Fujii:2015vha, Alexandru:2015xva, Hayata:2015lzj, Alexandru:2015sua, Alexandru:2016gsd, Alexandru:2016ejd}. 

The basic idea is to deform the integration contour $\mathcal{M}$ into steepest descent cycles inside its complexified space $\mathcal{M}_{\mathbb{C}}$ by using the Cauchy theorem when $F(\Phi)$ is holomorphic. We denote the holomorphic coordinate of $\mathcal{M}_{\mathbb{C}}$ as $\Phi=(z^1,\ldots,z^n)$, and the set of saddle points as 
\be
\Sigma=\{z_{\sigma}\}:=\left\{{\p F\over\p z^i}=0 \right\}. 
\ee
Using the K\"ahler metric on $\mathcal{M}_\mathbb{C}$, $\diff s^2=g_{i\overline{j}}\diff z^i\otimes \diff \overline{z}^{\overline{j}}$, we define the gradient flow by 
\be
{\diff z^i\over \diff t}=g^{i\overline{j}}\,\overline{\left({\p F(z)\over \p z^j}\right)}. 
\label{eq:gradient_flow_general}
\ee
As an important property of this differential equation, we have
\be
{\diff F\over \diff t}=\left|\p F\right|^2\ge 0. 
\ee
Therefore, along the flow line, the real part of the free energy increases while its imaginary part stays constant. This means that we can define the steepest descent and ascent cycles associated with each saddle point $z_{\sigma}$ by this gradient flow. Using solutions of the gradient flow $z(t)$, they are defined as 
\be
\mathcal{J}_{\sigma}=\{z(0)\, |\, z(t)\to z_{\sigma}, t\to -\infty\}, \quad 
\mathcal{K}_{\sigma}=\{z(0)\, |\, z(t)\to z_{\sigma}, t\to +\infty\}. 
\ee
$\mathcal{J}_{\sigma}$ and $\mathcal{K}_{\sigma}$ are called Lefschetz thimble and dual thimble, respectively. 
They are dual quantities in terms of the intersection pairing $\langle\cdot,\cdot \rangle$, i.e., $\langle \mathcal{J}_{\sigma},\mathcal{K}_{\tau}\rangle =\delta_{\sigma \tau}$, which means that one can decompose $\mathcal{M}$ in terms of $\mathcal{J}_{\sigma}$ as 
\be
\int_{\mathcal{M}} \diff \Phi \exp\left(-V F(\Phi)\right)=\sum_{\sigma\in \Sigma} \langle  \mathcal{M}, \mathcal{K}_{\sigma}\rangle \int_{\mathcal{J}_{\sigma}}\diff^n z \exp\left(-V F(z)\right). 
\ee
If all $\mathrm{Re}(F(z_{\sigma}))$ are different with each other in the limit $V\to \infty$, we replace the integral by the saddle-point approximation, and we obtain at the leading order that 
\be
Z=\sum_{\sigma} \langle  \mathcal{M}, \mathcal{K}_{\sigma}\rangle  \exp\left(-V F(z_{\sigma})\right). 
\label{eq:MFpartition_function_complex_saddles}
\ee

We can summarize the necessary steps of the mean-field approximation with the sign problem as follows: 
\begin{itemize}
\item Complexify the target space $\mathcal{M}$ to $\mathcal{M}_{\mathbb{C}}$, and find the saddle points $z_{\sigma}$ by solving the equation $\p F=0$ in $\mathcal{M}_{\mathbb{C}}$. 
\item Solve the gradient flow (\ref{eq:gradient_flow_general}), and construct Lefschetz thimbles $\mathcal{J}_{\sigma}$ and dual thimbles $\mathcal{K}_{\sigma}$.
\item Pick up the saddle point $z_{\sigma}$ that has the minimal free energy $\mathrm{Re}(F(z_{\sigma}))$ with nonzero intersection number $\langle\mathcal{M},\mathcal{K}_{\sigma}\rangle$.  
\end{itemize}

\subsection{Charge and complex conjugation for real-valued free energy}\label{sec:CK}

In Sec.~\ref{sec:complex_mean_field}, we explained the way to apply the Lefschetz thimble method to the mean-field calculations for the theory with the sign problem. 
However, since the free energy $F(z)$ is complex in general, it would not be clear if the mean-field free energy $F(z_{\sigma})$ becomes real in the above procedure. 
Following Ref.~\cite{Tanizaki:2015pua}, we explain that this is ensured under the charge conjugation of the original theory. 

To explain it, we consider a fermionic system and denote the chemical potential dependence of the free energy explicitly as $F=F(\Phi,\mu)$. 
In many examples of the fermionic system at finite densities, unbalance between the fermion and anti-fermion numbers due to the chemical potential causes the sign problem, and the free energy satisfies
\be
\overline{F(\Phi,\mu)}=F(\Phi,-\mu). 
\ee
In order to relate $F(\Phi,-\mu)$ to the original one, we consider the charge conjugation $C$, which flips the sign of the chemical potential. Then, we have 
\be
\overline{F(\Phi,\mu)}=F(C\cdot \Phi,\mu). 
\ee
Indeed, this is satisfied for Polyakov-loop extended Nambu--Jona-Lasinio model~\cite{Nishimura:2014rxa,Nishimura:2014kla, Fukushima:2003fw}, 
heavy-dense quantum chromodynamics (QCD)~\cite{Dumitru:2005ng, Fukushima:2006uv, 
Alexandrou:1998wv, Condella:1999bk, Alford:2001ug, Banks:1983me,
Pisarski:2000eq, Dumitru:2000in, Akerlund:2016myr, Hirakida:2016rqd, Bender:1992gn, Blum:1995cb}, perturbative QCD with finite $\mu$~\cite{Hands:2010zp, Reinosa:2015oua}, Thirring model at finite densities~\cite{Tanizaki:2015rda, Fujii:2015bua, Fujii:2015vha, Alexandru:2015xva, Hayata:2015lzj, Pawlowski:2013pje} and also QED$_2$ in Sec.~\ref{sec:2dimQED}. 
Even after complexification, the anti-linear extension of the charge conjugation plays an important role, 
\be
z\mapsto C\cdot \overline{z},
\ee
because 
\be
\overline{F(z,\mu)}=F(C\cdot \overline{z},\mu). 
\label{eq:CK_symmetry}
\ee
This property is called the CK-symmetry of the complexified theory in Refs.~\cite{Nishimura:2014rxa, Nishimura:2014kla}. 

Under a certain condition of the K\"ahler metric and the charge conjugation, one can show that $C\cdot \overline{z}(t)$ satisfies the same equation~(\ref{eq:gradient_flow_general}) for any solutions $z(t)$ of the gradient flow~\cite{Tanizaki:2015pua}. 
Now, one can classify the set of saddle points $z_{\sigma}$ into three cases, 
\be
\Sigma_\mathrm{CK}=\{z_{\sigma}=C\cdot \overline{z}_{\sigma}\},\quad \Sigma_{\pm}=\{\mathrm{Im}F(z_{\sigma})\gtrless 0\}. 
\ee
We here notice that $F(z_{\sigma})$ is real if the saddle-point is CK-invariant, $z_{\sigma}=C\cdot \overline{z}_{\sigma}$\footnote{If the theory has other discrete symmetries, such as parity, center symmetry, etc., then one can construct other CK-transformation by combining them with the charge and complex conjugation. In such cases, we call $z_{\sigma}\in \Sigma_{\mathrm{CK}}$ if $z_{\sigma}$ is invariant under one of them. }. 
If the mean-field theory is valid, then $F(z_{\sigma})$ of the selected saddle point must be real. 
Therefore, if one can show or assumes the validity of the mean-field approximation, it is enough to consider the gradient flow~(\ref{eq:gradient_flow_general}) inside the CK-invariant hyperplane, and one can pick up the CK-invariant saddle point of the minimal free energy with nonzero intersection number $\langle\mathcal{M},\mathcal{K}_{\sigma}\rangle$.  
We will see that this is the case for the sign problem of multi-flavor massless QED$_2$ in the zero-temperature limit. 

\section{Lefschetz-thimble calculus for multi-flavor massless QED$_2$}\label{sec:calculation}
In this section, we consider the phase diagram of multi-flavor massless QED$_2$ at finite $L$ and the zero-temperature limit $\beta\to \infty$. 
For this purpose, we compute the path integral (\ref{eq:partition_function_QED2}) based on the Lefschetz-thimble method, and the mean-field calculation with the complex saddle points given in Sec.~\ref{sec:formalism} turns out to be powerful for our purpose. 

\subsection{Complex saddle points of multi-flavor massless QED$_2$}
We use the same symbol for the complexified toron fields, $h_0$ and $h_1$. 
We define the K\"ahler metric on complex $(h_0,h_1)$ space $(\mathbb{C}/\mathbb{Z})^2$ by 
\be
\diff s^2 =\tau^2 \diff h_0 \otimes \diff \overline{h_0}+\diff h_1\otimes \diff \overline{h_1}. 
\ee
The gradient flow to be solved in this metric is given by 
\be
{\diff h_0\over \diff t}={1\over \tau^2}\overline{\left({\p F\over \p h_0}\right)}, \quad 
{\diff h_1\over \diff t}=\overline{\left({\p F\over \p h_1}\right)},  
\label{eq:gradient_flow_qed2}
\ee
where the explicit form of the right hand sides is given as
\bea
{\p F\over\p h_0}&=&\im \tau\sum_{a=1}^{N_f}\sum_{n=1}^{\infty}\Bigl\{\left(1+\mathrm{e}^{{2\pi\over \tau}  \left(n+h_1-1-\mu'_a\right)+2\pi \im h_0}\right)^{-1}- \left(1+\mathrm{e}^{{2\pi \over \tau} \left(n-h_1+\mu'_a\right)-2\pi \im h_0}\right)^{-1}\nonumber\\
&&+\left(1+\mathrm{e}^{{2\pi \over \tau} \left(n-h_1-\mu'_a\right)+2\pi \im h_0}\right)^{-1}- \left(1+\mathrm{e}^{{2\pi\over \tau}  \left(n+h_1-1+\mu'_a\right)-2\pi \im h_0}\right)^{-1}\Bigr\}, 
\label{eq:gradient0}\\
{\p F\over\p h_1}&=&2N_f\left(h_1-{1\over 2}\right)\nonumber\\
&&+\sum_{a=1}^{N_f}\sum_{n=1}^{\infty}\Bigl\{\left(1+\mathrm{e}^{{2\pi\over \tau}  \left(n+h_1-1-\mu'_a\right)+2\pi \im h_0}\right)^{-1}- \left(1+\mathrm{e}^{{2\pi \over \tau} \left(n-h_1+\mu'_a\right)-2\pi \im h_0}\right)^{-1}\nonumber\\
&&-\left(1+\mathrm{e}^{{2\pi \over \tau} \left(n-h_1-\mu'_a\right)+2\pi \im h_0}\right)^{-1}+ \left(1+\mathrm{e}^{{2\pi\over \tau}  \left(n+h_1-1+\mu'_a\right)-2\pi \im h_0}\right)^{-1}\Bigr\}. 
\label{eq:gradient1}
\eea
What we have to do for the Lefschetz-thimble method is to find the saddle points of $F$ in the complexified $(h_0,h_1)$ space. Therefore, from (\ref{eq:gradient0}) and (\ref{eq:gradient1}), we need to solve 
\bea
\sum_{a=1}^{N_f} \sum_{n=1}^{\infty}\Bigl\{\left(1+\mathrm{e}^{{2\pi\over \tau}  \left(n+h_1-1-\mu'_a \right)+2\pi \im h_0}\right)^{-1}- \left(1+\mathrm{e}^{{2\pi \over \tau} \left(n-h_1+\mu'_a \right)-2\pi \im h_0}\right)^{-1}\Bigr\}
&=&-N_f\left(h_1-{1\over 2}\right),\nonumber\\
\sum_{a=1}^{N_f}\sum_{n=1}^{\infty} \Bigl\{\left(1+\mathrm{e}^{{2\pi \over \tau} \left(n-h_1-\mu'_a \right)+2\pi \im h_0}\right)^{-1}-\left(1+\mathrm{e}^{{2\pi\over \tau}  \left(n+h_1-1+\mu'_a \right)-2\pi \im h_0}\right)^{-1}\Bigr\}
&=&N_f\left(h_1-{1\over 2}\right). \nonumber\\
\label{eq:saddle_point_condition_general}
\eea
By taking the zero-temperature limit $\tau\to 0$, one can simplify this saddle-point condition as follows. 
In the limit $\tau \to 0$, we notice that the summation over spatial momenta $n$ is represented by the floor function, 
\bea
&&\sum_{n=1}^{\infty}\Bigl\{\left(1+\mathrm{e}^{{2\pi\over \tau}  \left(n+h_1-1-\mu'_a \right)+2\pi \im h_0}\right)^{-1}- \left(1+\mathrm{e}^{{2\pi \over \tau} \left(n-h_1+\mu'_a \right)-2\pi \im h_0}\right)^{-1}\Bigr\}\nonumber\\
&\to&\mathrm{max}\left(0,\lfloor1-\mathrm{Re}(h_1)+\mu'_a+\tau\mathrm{Im}(h_0)\rfloor\right)-\mathrm{max}\left(0,\lfloor \mathrm{Re}(h_1)-\mu'_a-\tau\mathrm{Im}(h_0)\rfloor\right)\nonumber\\
&=&\lfloor 1-\mathrm{Re}(h_1)+\mu'_a+\tau\mathrm{Im}(h_0)\rfloor. 
\eea
Therefore, the saddle-point conditions (\ref{eq:saddle_point_condition_general}) at $\tau\ll 1$ must be well approximated by 
\bea
\sum_{a=1}^{N_f}\lfloor 1-\mathrm{Re}(h_1)+\mu'_a+\tau\mathrm{Im}(h_0)\rfloor&=&-N_f\left(h_1-{1\over 2}\right),\label{eq:saddlepoint_zeroT_0}\\
\sum_{a=1}^{N_f}\lfloor \mathrm{Re}(h_1)+\mu'_a+\tau\mathrm{Im}(h_0)\rfloor&=&N_f\left(h_1-{1\over 2}\right). 
\label{eq:saddlepoint_zeroT_1}
\eea
Since the left hand sides of these equations are real and integers because they are left- and right-handed fermion numbers, so are the right hand sides. This claims that $h_1$ must be real and, especially\footnote{If some arguments of the floor function converge to integers in $\tau\to 0$, complex saddle points with other $h_1$'s can exist. We shall find them by solving the gradient flow numerically inside the CK-invariant plane in Sec.~\ref{sec:gradient_flow}. However, all of them turns out not to be selected as correct mean fields, since they do not have the minimal free energy with non-zero intersection number, and it is reasonable because the fermion numbers at those saddle points are not quantized even at the zero-temperature limit. Therefore, we neglect such subtle possibilities for a while. },
\be
h_1\in \left({1\over 2}+{1\over N_f}\mathbb{Z}\right)/\mathbb{Z}. 
\label{eq:saddle_condition_h1}
\ee
For further consideration on the saddle points, we need to specify the number of fermion flavors $N_f$. We will consider the case for $N_f=1$, $2$, and $3$ in following subsections. 

For the mean-field calculations, the computation of the free energy is also an important ingredient. 
In the limit $\tau\to 0$, one can find after a slightly technical computation that the free energy (\ref{eq:free_energy_qed2}) also accepts the much simplified expression as 
\bea
F&=& N_f\left(h_1-{1\over 2}\right)^2+{1\over 2}\sum_{a=1}^{N_f}\Bigl\{\lfloor 1-h_1+\mu'_a+\tau \mathrm{Im}(h_0)\rfloor^2+\lfloor h_1+\mu'_a+\tau \mathrm{Im}(h_0)\rfloor^2\Bigr\}\nonumber\\
&&+\left(h_1-{1\over 2}\right)\sum_{a=1}^{N_f}\Bigl(\lfloor 1-h_1+\mu'_a+\tau \mathrm{Im}(h_0)\rfloor-\lfloor h_1+\mu'_a+\tau \mathrm{Im}(h_0)\rfloor\Bigr)\nonumber\\
&&+\sum_{a=1}^{N_f}(\tau \im h_0-\mu'_a)\Bigl(\lfloor 1-h_1+\mu'_a+\tau \mathrm{Im}(h_0)\rfloor+\lfloor h_1+\mu'_a+\tau \mathrm{Im}(h_0)\rfloor\Bigr). 
\eea
One should notice that $h_0$ dependence totally disappears in the vicinity of the complex saddle points satisfying (\ref{eq:saddlepoint_zeroT_0}) and (\ref{eq:saddlepoint_zeroT_1}). 
This is due to the charge neutrality condition on the torus;
\bea
\sum_{a=1}^{N_f}\lfloor 1-h_1+\mu'_a+\tau\mathrm{Im}(h_0)\rfloor
+\sum_{a=1}^{N_f}\lfloor h_1+\mu'_a+\tau\mathrm{Im}(h_0)\rfloor=0.
\label{eq:charge_neutrality}
\eea
Then, we find that the imaginary part of the constrained free energy also disappears under that condition. This fact is remarkable because use of complex saddle points weakens the sign problem of the massless multi-flavor QED$_2$. 

\subsection{1-flavor case}
Let us first consider the case $N_f=1$. The saddle point condition for $h_1$, (\ref{eq:saddle_condition_h1}), says that $h_1={1\over 2}$. Then, Eqs.~(\ref{eq:saddlepoint_zeroT_0}) and (\ref{eq:saddlepoint_zeroT_1}) reduce to the single equation, 
\be
\left\lfloor {1\over 2}+\mu'+\tau\mathrm{Im}(h_0)\right\rfloor=0,
\ee
which can be trivially satisfied by choosing $\tau \mathrm{Im}(h_0)=-\mu'$. This saddle-point condition says that both the left- and right-handed fermion densities are always zero for any chemical potential. 

One can understand this result in an easier way. The chemical potential dependence of the free energy (\ref{eq:free_energy_qed2})  can be written as 
\be
F(h_0,h_1,\mu)=F\left(h_0+{\im\over \tau} \mu',h_1\right). 
\ee
Therefore, just by shifting $h_0$ as $h_0-\im \mu'/\tau$, the free energy becomes a real function, and the sign problem disappears. The partition function does not change under this continuous change of contours, so the chemical potential dependence does not appear in the single-flavor case.

\subsection{2-flavor case}
\begin{figure}[t]\centering
\begin{minipage}{.45\textwidth}
\subfloat[Mean-field free energy]{
\includegraphics[scale=0.55]{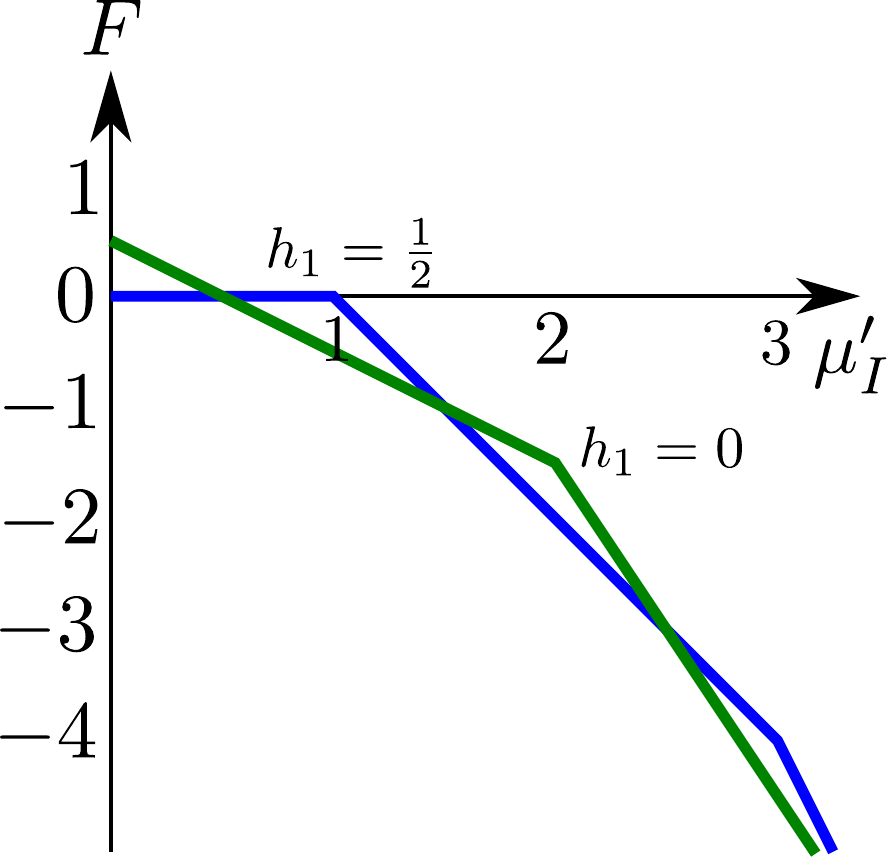}
\label{fig:2flavorSchwinger_1}
}\end{minipage}
\begin{minipage}{.45\textwidth}
\subfloat[Fermion number]{
\includegraphics[scale=0.62]{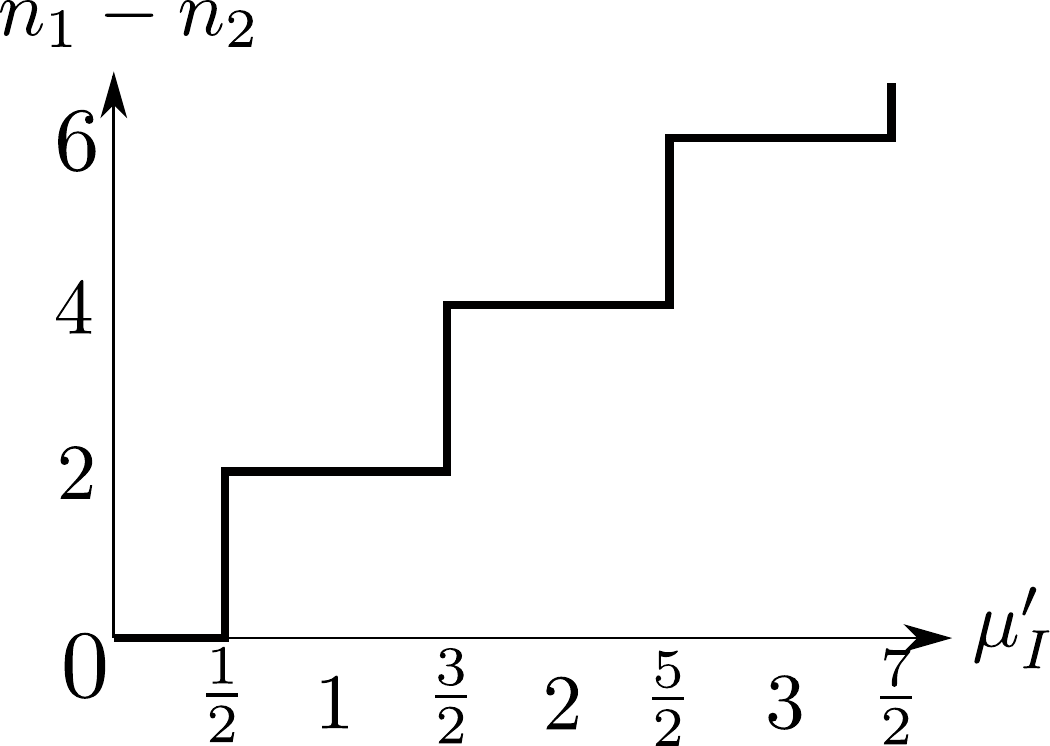}
\label{fig:2flavorSchwinger_2}
}\end{minipage}
\caption{(a)~Mean-field free energy at complex saddles with $h_1={0}$ (green line) and $h_1={1\over 2}$ (blue line) as a function of $\mu'_I=\mu'_1-\mu'_2$. (b)~Behaviors of the difference of fermion numbers at $\tau=0$ as a function of $\mu'_I$.  }
\label{fig:2flavorSchwinger}
\end{figure}

Let us next consider the case $N_f=2$.
  The saddle-point condition (\ref{eq:saddle_condition_h1}) requires that $h_1={1\over 2}$ or $h_1=0$. For $h_1={0}$,  Eqs.~(\ref{eq:saddlepoint_zeroT_0}) and (\ref{eq:saddlepoint_zeroT_1}) become 
\bea
\left\lfloor 1+\mu'_1+\tau\mathrm{Im}(h_0)\right\rfloor+\left\lfloor 1+\mu'_2+\tau\mathrm{Im}(h_0)\right\rfloor&=&1,\\ 
\left\lfloor \mu'_1+\tau\mathrm{Im}(h_0)\right\rfloor+\left\lfloor \mu'_2+\tau\mathrm{Im}(h_0)\right\rfloor&=&-1. 
\eea
Since these two conditions are equivalent, and they can be satisfied by setting $\tau\mathrm{Im}(h_0)=-{1\over 2}(\mu'_1+\mu'_2)$ because $\lfloor x \rfloor +\lfloor -x\rfloor=-1$ for any $x\in \mathbb{R}$. 
Introducing the dimensionless isospin chemical potential by 
\be
\mu'_I=\mu'_1-\mu'_2, 
\ee
the free energy of this state is 
\be
F\left(-{\im\over 2\tau}(\mu'_1+\mu'_2),0\right)={1\over 2}+2\left\lfloor{\mu'_I\over 2}\right\rfloor+2\left\lfloor{\mu'_I\over 2}\right\rfloor^2-\mu'_I\left(1+2\left\lfloor{\mu'_I\over 2}\right\rfloor\right). 
\label{eq:2flavor_free_energy_0}
\ee
For $h_0={1\over 2}$, Eqs.~(\ref{eq:saddlepoint_zeroT_0}) and (\ref{eq:saddlepoint_zeroT_1}) become the single equation,
\be
\left\lfloor {1\over 2}+\mu'_1+\tau\mathrm{Im}(h_0)\right\rfloor+\left\lfloor {1\over 2}+\mu'_2+\tau\mathrm{Im}(h_0)\right\rfloor=0. 
\ee
Again, this is satisfied by setting $\tau\mathrm{Im}(h_0)=-{1\over 2}(\mu'_1+\mu'_2)$, and the free energy becomes 
\be
F\left(-{\im\over 2\tau}(\mu'_1+\mu'_2),{1\over 2}\right)=2\left\lfloor{1\over 2}+{\mu'_I\over 2}\right\rfloor^2-2\mu'_I\left\lfloor{1\over 2}+{\mu'_I\over 2}\right\rfloor. 
\label{eq:2flavor_free_energy_1}
\ee
Soon later, we shall show that both complex saddles have nonzero intersection number, and thus we must pick up a state with the lower free energy at a given chemical potential to determine the phase diagram. 
Figure~\ref{fig:2flavorSchwinger_1} shows (\ref{eq:2flavor_free_energy_0}) and (\ref{eq:2flavor_free_energy_1}) with green and blue solid lines, respectively, and we can readily find that the phase transition at $\tau=0$ happens when $\mu'_I$ is a half integer. 
The difference of the fermion numbers, $n_1-n_2=-2{\p\over \p\mu'_I}F$, is shown in Fig.~\ref{fig:2flavorSchwinger_2}. 
This result is consistent with the exact computation done in Ref.~\cite{Narayanan:2012qf} and also with the result by the matrix product states in Ref.~\cite{Banuls:2016gid}. 

To complete our argument on the $2$-flavor case, let us discuss the intersection numbers at the saddle points with $h_1=0$ and $h_1={1\over 2}$. It is important to notice that the saddle-point conditions for both cases are solved by $\tau \mathrm{Im}(h_0)=-{1\over 2}(\mu'_1+\mu'_2)$. Indeed, the free energy $F(h_0,h_1)$ becomes a real-valued function after the shift 
\be
h_0\mapsto h_0-\im {\mu'_1+\mu'_2\over 2}. 
\ee
The above two saddle points at $h_1=0,\,{1\over 2}$ correspond to local minimum of the real-valued free energy after this shift of $h_0$, and thus the usual mean-field approximation becomes valid. 
This means that the intersection numbers of both saddle points are equal to $1$:
\be
\left(\mathbb{R}/\mathbb{Z}-\im{\mu'_1+\mu'_2 \over 2}\right)\times (\mathbb{R}/\mathbb{Z})=\mathcal{J}_{\left(-\im{\mu'_1+\mu'_2 \over 2},0\right)}+\mathcal{J}_{\left(-\im{\mu'_1+\mu'_2 \over 2},{1\over 2}\right)}, 
\ee
when neglecting the consistently subdominant thimbles. 
This argument, however, also shows that the sign problem of two-flavor QED$_2$ can be eliminated just by the constant shift of integration variables. 

\subsection{3-flavor case}\label{sec:saddles_3flavor}
Let us consider the case $N_f=3$ as a last example.
This is the first nontrivial case, in which we cannot eliminate the sign problem by simply shifting the $h_0$ field. 
Without loss of generality, we can set $\mu_3=0$ by shift of $h_0$. In the following, we solve the problem under this condition. The saddle-point condition for $h_1$, (\ref{eq:saddle_condition_h1}) says that 
\be
h_1={1\over 2},\, {1\over 2}\pm {1\over 3}. 
\label{eq:3flavor_h1_condition}
\ee
We solve the saddle-point conditions for $h_0$, (\ref{eq:saddlepoint_zeroT_0}) and (\ref{eq:saddlepoint_zeroT_1}), by separating cases. 
\begin{enumerate}
\item We set $h_1={1\over 2}$. Introducing $y=\tau \mathrm{Im}(h_0)+{1\over 2}$, the saddle-point condition at $T=0$, (\ref{eq:saddlepoint_zeroT_0}) and (\ref{eq:saddlepoint_zeroT_1}), can be written as 
\be
\lfloor \mu'_1+y\rfloor +\lfloor \mu'_2+y\rfloor +\lfloor y\rfloor =0. 
\ee
To solve this condition, we set, for integers $n$ and $m$,
\be
\lfloor \mu'_1+y\rfloor=n, \, \lfloor \mu'_2+y\rfloor=m,\, \lfloor y\rfloor =-n-m. 
\ee
The region is surrounded by hexagon, which is the convex hull of the set of six points,
\be
\left\{
\begin{array}{c}
(2n+m-1, n+2m-1),(2n+m,n+2m-1),(2n+m+1,n+2m),\\(2n+m+1,n+2m+1), (2n+m, n+2m+1), (2n+m-1, n+2m)
\end{array}\right\}. 
\ee
In Fig.~\ref{fig:3flavorSchwinger_1}, we show the structure of these hexagons with corresponding fermion numbers $(n_1,n_2)=(2n,2m)$. 
The free energy for this case becomes 
\bea
F\left(h_0,{1\over 2}\right)&=&n^2+m^2+(-n-m)^2-2n \mu'_1-2m \mu'_2-2(-n-m)\mu'_3. 
\eea
Here, we reinstate $\mu'_3$ just for keep the expression symmetric. 

\begin{figure}[t]
\centering
\begin{minipage}{.45\textwidth}
\subfloat[$h_1={1\over 2}$]{
\includegraphics[scale=0.5]{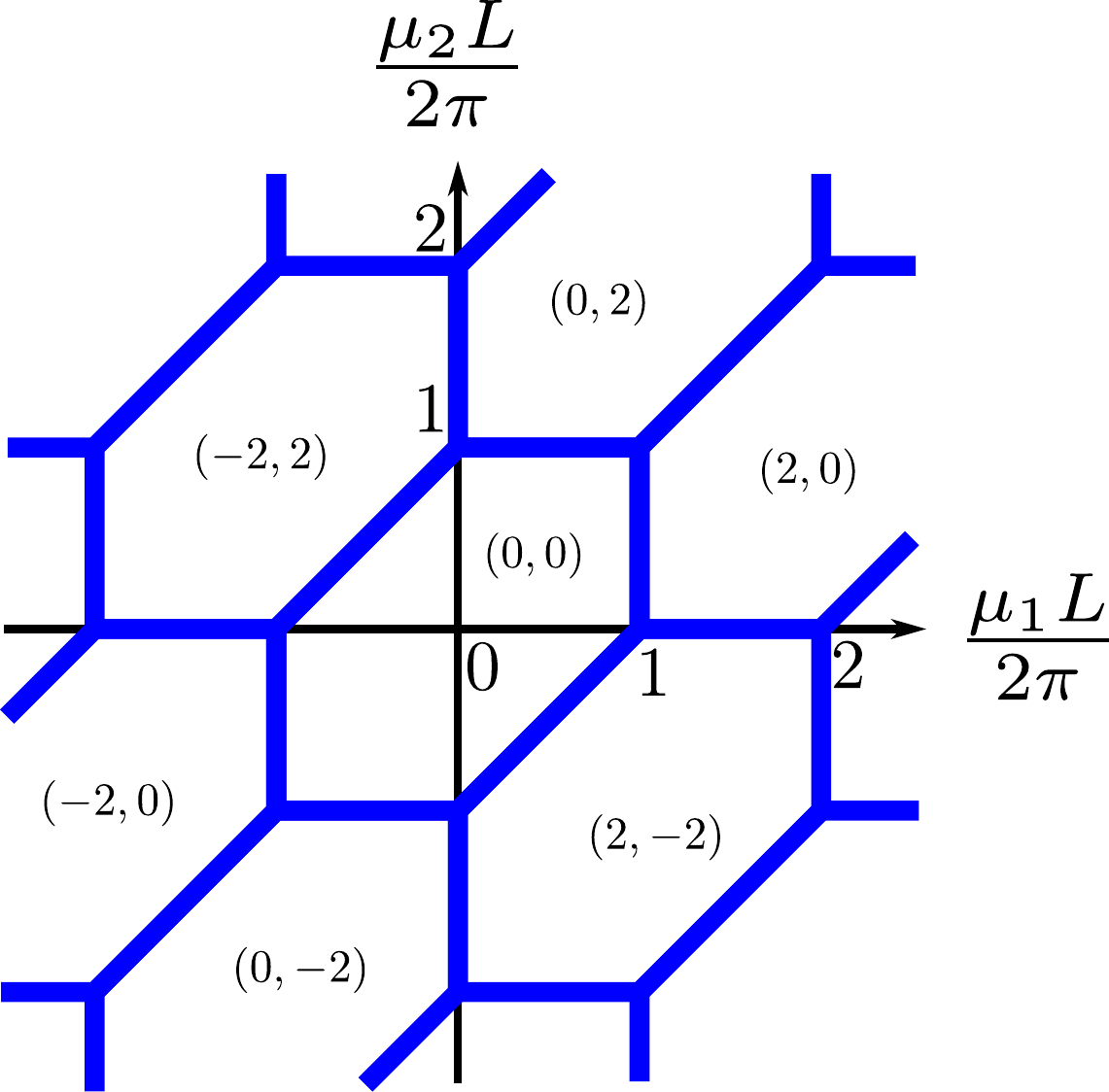}
\label{fig:3flavorSchwinger_1}
}\end{minipage}
\begin{minipage}{.45\textwidth}
\subfloat[$h_1={1\over 2}\pm {1\over 3}$]{
\includegraphics[scale=0.32]{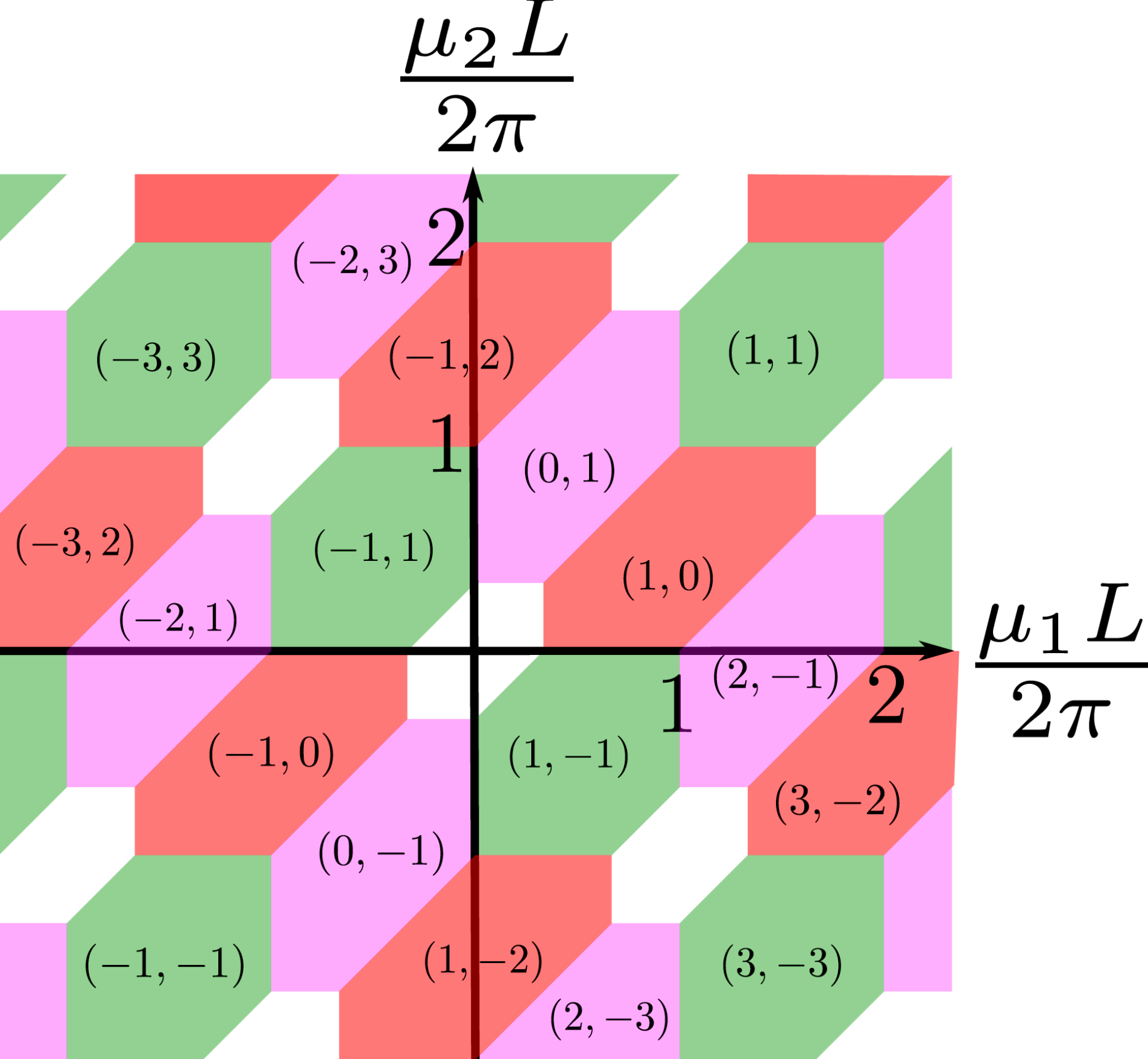}
\label{fig:3flavorSchwinger_2}
}\end{minipage}
\caption{Classification of complex saddle points for the three-flavor case with $\mu_3=0$. In each figure, we set $h_1={1\over 2}$ and $h_1={1\over 6},{5\over 6}$, respectively. We show the fermion numbers $(n_1,n_2)$ in each region, and $n_3=-n_1-n_2$ due to the charge neutrality. }
\label{fig:3flavorSchwinger_candidates}
\end{figure}

\item We set $h_1={1\over 2}\pm {1\over 3}$. Setting $y=\tau\mathrm{Im}(h_0)+{1\over 6}$, the conditions  (\ref{eq:saddlepoint_zeroT_0}) and (\ref{eq:saddlepoint_zeroT_1}) become 
\bea
\sum_{a=1}^{3} \left\lfloor \mu'_a+y\right\rfloor
=-1,\quad 
\sum_{a=1}^{3} \left\lfloor \mu'_a+{2\over 3}+y\right\rfloor
=1. 
\eea
Setting $\mu'_3=0$, we solve this condition by separating it into three cases. 
\begin{enumerate}
\item \label{case:n1odd}
We first consider the case 
\be
\left\lfloor \mu'_1+{2\over 3}+y\right\rfloor=\left\lfloor \mu'_1+y\right\rfloor+1=n+1,\; \left\lfloor \mu'_2+{2\over 3}+y\right\rfloor=\left\lfloor \mu'_2+y\right\rfloor=m. 
\ee
The corresponding region becomes the hexagon, which is the convex hull of 
\be\hspace{-2.5em}
\left\{
\begin{array}{c}
(2n+m+{1\over 3}, n+2m),(2n+m+1,n+2m),(2n+m+{5\over 3},n+2m+{2\over 3}),\\(2n+m+{5\over 3},n+2m+1), (2n+m+1, n+2m+1), (2n+m+{1\over 3}, n+2m+{1\over 3})
\end{array}\right\}. 
\ee
\item \label{case:n2odd}
Next, consider the case 
\be
\left\lfloor \mu'_1+{2\over 3}+y\right\rfloor=\left\lfloor \mu'_1+y\right\rfloor=n,\; \left\lfloor \mu'_2+{2\over 3}+y\right\rfloor=\left\lfloor \mu'_2+y\right\rfloor+1=m+1. 
\ee
The corresponding region becomes the hexagon, which is the convex hull of 
\be\hspace{-2.5em}
\left\{
\begin{array}{c}
(2n+m, n+2m+{1\over 3}),(2n+m+{1\over 3},n+2m+{1\over 3}),(2n+m+{1},n+2m+{1\over 3}),\\(2n+m+1,n+2m+{5\over 3}), (2n+m+{2\over 3}, n+2m+{5\over 3}), (2n+m, n+2m+{1})
\end{array}\right\}. 
\ee
\item \label{case:n1oddn2odd}
Thirdly, consider the case 
\be
\left\lfloor \mu'_1+{2\over 3}+y\right\rfloor=\left\lfloor \mu'_1+y\right\rfloor+1=n+1,\; \left\lfloor \mu'_2+{2\over 3}+y\right\rfloor=\left\lfloor \mu'_2+y\right\rfloor+1=m+1. 
\ee
The corresponding region becomes the hexagon, which is the convex hull of 
\be\hspace{-2.5em}
\left\{
\begin{array}{c}
(2n+m+1, n+2m+1),(2n+m+{5\over 3},n+2m+1),(2n+m+2,n+2m+{4\over 3}),\\(2n+m+2,n+2m+2), (2n+m+{4\over 3}, n+2m+2), (2n+m+1, n+2m+{5\over 3})
\end{array}\right\}. 
\ee
\end{enumerate}
The result for $h_1={1\over 2}\pm {1\over 3}$ is summarized in Fig.~\ref{fig:3flavorSchwinger_2}, where orange, pink and green regions correspond to the above cases \ref{case:n1odd}, \ref{case:n2odd} and \ref{case:n1oddn2odd}, respectively. The blank region of Fig.~\ref{fig:3flavorSchwinger_2} means that there is no corresponding saddle point with $h_1={1\over 2}\pm {1\over 3}$ at those chemical potentials. 
The free energy becomes
\bea
F \left(h_0, \frac{1}{2}\pm \frac{1}{3}\right)
&=&-{1\over 3}+{1\over 2}\sum_{a=1}^{3}\left( \left\lfloor {5\over 6}+\mu'_a+\tau \mathrm{Im}(h_0)\right\rfloor^2+\left\lfloor {1\over 6}+\mu'_a+\tau \mathrm{Im}(h_0)\right\rfloor^2\right)\nonumber\\
&&-\sum_{a=1}^{3}\mu'_a \left( \left\lfloor {5\over 6}+\mu'_a+\tau \mathrm{Im}(h_0)\right\rfloor+\left\lfloor {1\over 6}+\mu'_a+\tau \mathrm{Im}(h_0)\right\rfloor\right). 
\eea
\end{enumerate}

\begin{figure}[t]
\centering
\includegraphics[scale=0.32]{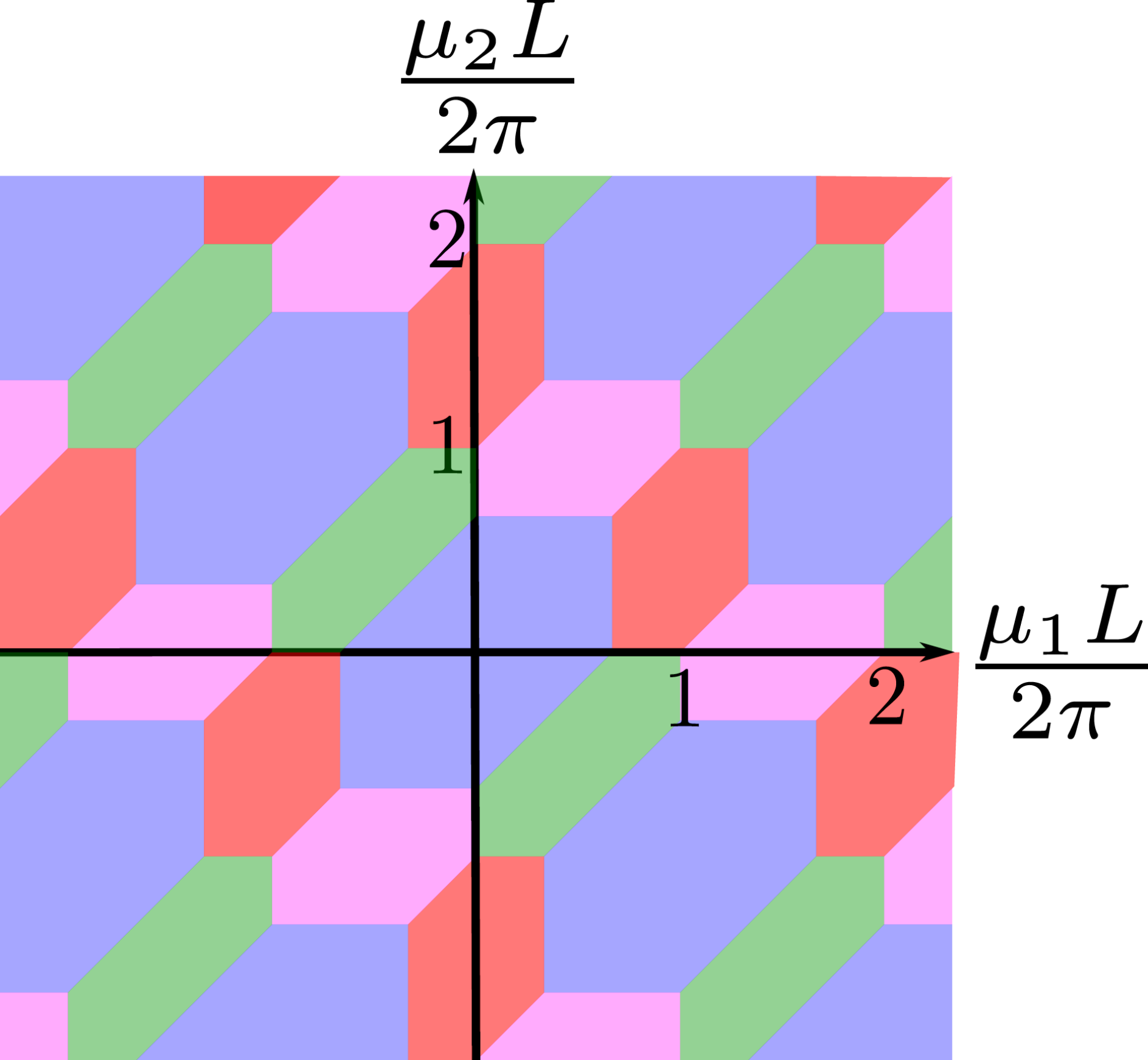}
\caption{Phase boundaries of the three-flavor massless QED$_2$ at $T=0$ and $\mu_3=0$. Inside blue hexagons, the phases in Fig.~\ref{fig:3flavorSchwinger_1} are selected, i.e., $h_1={1\over 2}$. Hexagons with other colors represent the phase with $h_1={1\over 2}\pm {1\over 3}$, and the same color with Fig.~\ref{fig:3flavorSchwinger_2} is used. }
\label{fig:3flavorSchwinger}
\end{figure}

We now have solved all the possible complex saddle points, and the corresponding free energies are computed. 
Since all the free energies are real and independent of $\tau$, the mean-field calculation with those complex saddle points becomes exact in the zero-temperature limit $\tau\to 0$. 
As we have explained in Sec.~\ref{sec:formalism}, we need to compute the gradient flow (\ref{eq:gradient_flow_qed2}) to construct the dual thimbles for this method, and select the phase with minimal free energy with nonzero intersection number. 
To streamline the discussion, however, we postpone the computation of the gradient flow to Sec.~\ref{sec:gradient_flow}, and we \textit{temporarily} assume that all the above complex saddle points contribute to the partition function in the formula (\ref{eq:MFpartition_function_complex_saddles}). 
This temporal assumption shall be verified by solving the gradient flow in the next section. 

We can readily compute the phase boundaries of the three-flavor massless QED$_2$ by using the information about complex saddle points. 
If there are several complex solutions at a given chemical potential by comparing Figs.~\ref{fig:3flavorSchwinger_1} and \ref{fig:3flavorSchwinger_2}, we select the phase with the lower free energy. 
As a result, we obtain the phase boundaries given in Fig.~\ref{fig:3flavorSchwinger}. 
One can explicitly compare this result with the exact computation done by Ref.~\cite{Lohmayer:2013eka} to find its correctness (see Figure 4 of Ref.~\cite{Lohmayer:2013eka}). 

\section{Gradient flow inside CK-invariant space}\label{sec:gradient_flow}
In order to check the intersection numbers of the dual thimbles, we solve the gradient flow numerically for three-flavor massless QED$_2$ in this section. 
We first emphasize the technical importance of the CK-transformation in order to visualize the gradient flow, and apply this technique to numerically solve the flow for $N_f=3$. 
All the following procedures are valid also for larger $N_f$'s. 

The big technical issue of the mean-field approximation with complex saddles is that we have to compute the intersection numbers between $(\mathbb{R}/\mathbb{Z})^2$ and dual thimbles $\mathcal{K}_{\sigma}$ inside the complexified space $(\mathbb{C}/\mathbb{Z})^2$. 
Since it is impossible to draw figures in the four-dimensional manifold, we do not know a manifestly clear way to compute such quantities. 
The CK-transformation plays an important role to manage this issue for the massless QED$_2$. 

Since the original theory has the symmetry under charge conjugation, one can construct the CK symmetry discussed in Sec.~\ref{sec:CK}~\cite{Tanizaki:2015pua, Nishimura:2014rxa, Nishimura:2014kla}, 
\be
\overline{F(h_0,h_1,\mu)}=F(-\overline{h_0},\overline{h_1},\mu). 
\ee
Especially if $\mathrm{Re}(h_0)=0$ and $\mathrm{Im}(h_1)=0$, every physical quantity becomes real. One can also notice that the saddle points obtained in the previous section satisfy this restriction. Therefore, in the following, we solve the gradient flow under the condition 
\be
h_0=\im y/\tau,\; h_1=x, 
\ee
with real $x$ and $y$. Indeed, if one starts the gradient flow from a generic point inside the CK-invariant plane, then the flow line lies inside the CK-invariant plane~\cite{Tanizaki:2015pua}. The gradient flow inside the CK-invariant plane obeys 
\bea
{\diff x\over \diff t}&=&2N_f\left(x-{1\over 2}\right)\nonumber\\
&&+\sum_{a=1}^{N_f}\sum_{n=1}^{\infty}\Bigl\{\left(1+\mathrm{e}^{{2\pi\over \tau}  \left(n+x-1-\mu'_a- y\right)}\right)^{-1}- \left(1+\mathrm{e}^{{2\pi \over \tau} \left(n-x+\mu'_a+ y\right)}\right)^{-1}\nonumber\\
&&-\left(1+\mathrm{e}^{{2\pi \over \tau} \left(n-x-\mu'_a- y\right)}\right)^{-1}+ \left(1+\mathrm{e}^{{2\pi\over \tau}  \left(n+x-1+\mu'_a+ y\right)}\right)^{-1}\Bigr\}, \\
{\diff y\over \diff t}&=&-\sum_{a=1}^{N_f}\sum_{n=1}^{\infty}\Bigl\{\left(1+\mathrm{e}^{{2\pi\over \tau}  \left(n+x-1-\mu'_a- y\right)}\right)^{-1}- \left(1+\mathrm{e}^{{2\pi \over \tau} \left(n-x+\mu'_a+ y\right)}\right)^{-1}\nonumber\\
&&+\left(1+\mathrm{e}^{{2\pi \over \tau} \left(n-x-\mu'_a- y\right)}\right)^{-1}- \left(1+\mathrm{e}^{{2\pi\over \tau}  \left(n+x-1+\mu'_a+ y\right)}\right)^{-1}\Bigr\}. 
\eea
We can now visualize the gradient flow inside the CK-invariant plane to compute the nontrivial quantities, intersection numbers. 
We numerically solve this equation to find out the structure of the gradient flow for the case $N_f=3$, and judge the contribution to the free energy from complex saddle points. 

\begin{figure}[t]
\centering
\includegraphics[scale=0.5]{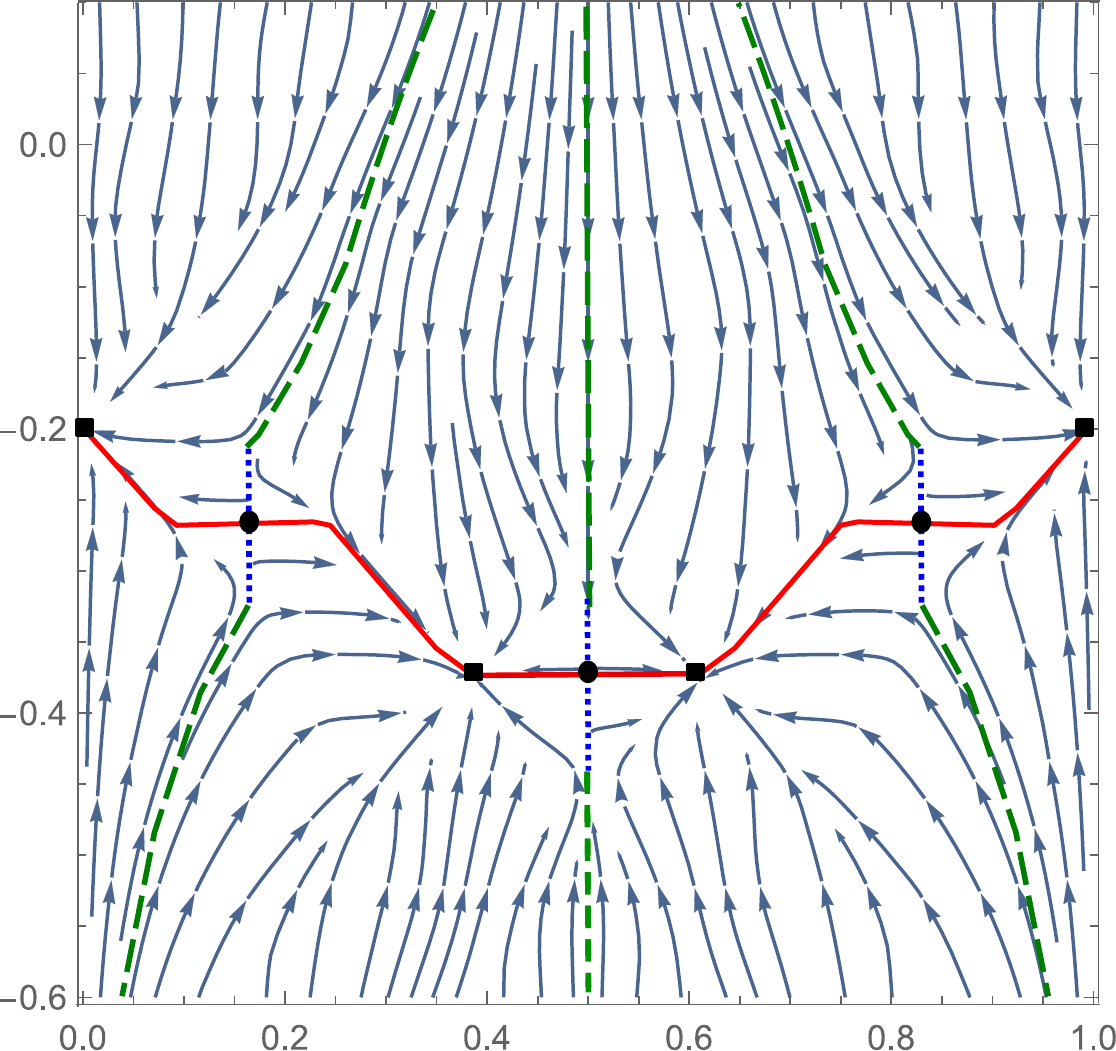}
\caption{Gradient flow in the CK-invariant plane $\mathrm{Re}(h_1)$-$\tau \mathrm{Im}(h_0)$ at $(\mu'_1,\mu'_2)=(0.75,0.2)$ and $\tau=0.1$, and the schematic illustration of Lefschetz thimbles. Black blobs show the saddle points in the previous section, and black squares show other subdominant saddles. 
Blue doted lines show the set of approximate saddle points. Red solid lines are Lefschetz thimbles $\mathcal{J}_{\sigma}$, and green dashed lines are parts of dual thimbles $\mathcal{K}_{\sigma}$ connecting to the line segment of approximate saddles. }
\label{fig:3flavorSchwinger_flow}
\end{figure}

We show the gradient flow and the schematic illustration of Lefschetz and dual thimbles in Fig.~\ref{fig:3flavorSchwinger_flow} at $(\mu'_1,\mu'_2)=(0.75,0.2)$. 
The horizontal axis represents $x=\mathrm{Re}(h_1)$, and the vertical axis does $y=\tau\mathrm{Im}(h_0)$. 
We can find black blobs at $h_1={1\over 2},{1\over 2}\pm{1\over 3}$, and they are saddle points computed in Sec.~\ref{sec:saddles_3flavor}. 
There are blue dotted line segments through those saddles, which consist of approximate saddles at $\tau=0.1$. 
This is because the saddle-point condition at $\tau=0$, (\ref{eq:saddlepoint_zeroT_0}) and (\ref{eq:saddlepoint_zeroT_1}), does not designate the specific value of $\tau\mathrm{Im}(h_0)$. At sufficiently low temperatures, the set of approximate saddle points form a line segment. 
The line segment of approximate saddles separate gradient flows, which moves almost horizontally towards right and left, and the flow lines emanating from it form a diamond-like shape. 
Black squares show subdominant saddles, at which $h_1$ is not appropriately quantized as (\ref{eq:3flavor_h1_condition}), and they are irrelevant in the mean-field calculation. 


We can now comment on when the sign problem of the reweighting is severe. Only inside the diamond-shaped regions, overall number density vanishes, $\sum_{a=1}^{3}n_a=0$, and the sign problem along the $h_0$-integration becomes absent. 
Therefore, if the real axis is away from the diamond-shaped region of the dominant saddle in Fig.~\ref{fig:3flavorSchwinger_flow}, the conventional reweighting suffers from the severe sign problem. 
We can solve this problem by deforming the integration contours to Lefschetz thimbles.

The existence of the line segment of approximate saddles, however, makes it difficult to solve the gradient flow with good accuracy especially around the saddle points. This brings us the numerical problem to compute the Lefschetz thimbles $\mathcal{J}_{\sigma}$ and dual thimbles $\mathcal{K}_{\sigma}$. 
However, what we need for the mean-field calculation with complex saddle points is the intersection number between the original integration cycle and dual thimbles $\mathcal{K}_{\sigma}$. Since the intersection number is a topological quantity, we do not need to compute $\mathcal{K}_{\sigma}$ exactly, and we can approximate $\mathcal{K}_{\sigma}$ as the line segment that contains $z_{\sigma}$ and the flow sucked into the edges of the line segment. 
In Fig.~\ref{fig:3flavorSchwinger_flow}, we show this approximate Lefschetz thimbles $\mathcal{J}_{\sigma}$ by red solid lines, and the approximate dual thimbles $\mathcal{K}_{\sigma}$ by green dashed lines combined with blue dotted lines. 
That is, if the line segment or the flow line sucked into the line segment intersects with the real axis, we must take the complex saddle into the Lefschetz thimble decomposition of the partition function. 


This confirms that all the saddle points discussed in Sec.~\ref{sec:calculation} contribute, and it justifies the mean-field calculation done in the previous section. Therefore, the phase boundaries for $N_f=3$ given in Fig.~\ref{fig:3flavorSchwinger} are exact at the zero-temperature $\tau=0$. 

\section{Conclusion and perspective}
\label{sec:conclusion}

We show that the mean-field approximation with complex saddle points gives the rigorous result on the phase structure of multi-flavor massless QED$_2$ at zero temperature and finite densities. 
We derive the saddle-point condition and concretely solve it for $N_f=1$,~$2$,~and~$3$ cases. 
By computing the gradient flow, we check the intersection number to list up the candidates of complex mean fields, and draw the phase diagram by comparing the mean-field free energies with nonzero intersection numbers. 

Multi-flavor massless QED$_2$ suffers from the sign problem, and the conventional reweighting technique does not work for general large chemical potentials. 
We have shown that the Lefschetz-thimble method completely solves this problem at $T=0$ for all the chemical potentials: After deforming the integration path into the complex space using the gradient flow, we can use the saddle-point method and the reweighting factor becomes consistent with $1$. 
This means that the lattice Monte Carlo simulation on Lefschetz thimbles solves the exponential complexity of multi-flavor massless QED$_2$ at finite densities. 

We would like to emphasize that this is highly nontrivial result. Indeed, there is an example of the fermion sign problem, the one-site Hubbard model, where the exponential complexity does remain even after the deformation of the integration path with the gradient flow~\cite{Tanizaki:2015rda}. 
In that example, complex zeros of the fermion determinant separate the original integration cycle into multiple Lefschetz thimbles, and the summing up them gives an exponentially small reweighting factor in terms of the inverse temperature $\beta$. 
Both one-site Hubbard model and multi-flavor massless QED$_2$ are strongly-coupled fermionic systems, and they have the first-order phase transition induced by jumps of the fermion number. 
Why do these examples have the big difference in the Lefschetz-thimble approach?

In the one-site Hubbard model~\cite{Tanizaki:2015rda}, there are infinitely many Lefschetz thimbles that contribute in the zero-temperature limit. Integration on each Lefschetz thimble at the saddle point $z_{\sigma}$ is represented by $\exp(-\beta F(z_{\sigma}))$ in our notation, and the difference of the free energy $\beta(F(z_{\sigma})-F(z_{\tau}))$ remains finite in the limit $\beta\to \infty$. This means that each Lefschetz-thimble integral gives a comparable contribution, and the interference among them affects significantly in the one-site Hubbard model. 
On the other hand, $F(z_{\sigma})$'s in the multi-flavor massless QED$_2$ are independent of $\beta$, and thus $\beta(F(z_{\sigma})-F(z_{\tau}))\to \pm \infty$ in the limit $\beta\to \infty$. One of the Lefschetz thimbles dominate the others, and the destructive interference does not occur in massless QED$_2$. 

Another speculative remark is that the charge neutrality condition (\ref{eq:charge_neutrality})
at the saddle point plays an important role to eliminate the sign problem in multi-flavor massless QED$_2$. 
In this model, the unbalance of oppositely charged particles creates the sign problem when integrating over the phase of the temporal Wilson loop, $h_0$, because of the violation of charge neutrality at each real configuration. 
Use of the complexified fields makes the charge neutrality manifest, and the $h_0$ dependence of the free energy totally disappears in the vicinity of the complex saddle points. 
It is an important future study to understand what is a consequence of the color neutrality condition of other QCD-like gauge theories. 

As a final remark, let us comment on the limitation of this study when we discuss the QCD sign problem. 
One of the biggest difference between massless QED$_2$ and the chiral limit of QCD is the property of the chiral symmetry. 
In QED$_2$, the chiral symmetry is broken by the axial anomaly, and the massless Nambu--Goldstone boson does not appear. Indeed, the spontaneous symmetry breaking of the continuous symmetry is forbidden in $(1+1)$-dimensional field theories~\cite{Coleman:1973ci, mermin1966absence}. 
On the other hand, QCD has the spontaneous chiral symmetry breaking, and massless pions exist. The existence of charged massless pions gives the big obstacle when applying the conventional reweighting technique to the sign problem of QCD at the zero temperature and finite density, which causes the early-onset problem of baryon number densities at half of the pion mass~\cite{Gibbs:1986ut, Gibbs:1986hi, Barbour:1997bh, Barbour:1997ej, Stephanov:1996ki, Cohen:2003kd, Splittorff:2006fu, Splittorff:2007ck, Cherman:2010jj, Hidaka:2011jj, Nagata:2012tc}. 
It is not yet known whether the massless pions bring any difficulties to the Lefschetz-thimble approach, and it cannot be discussed with massless QED$_2$ because of the different origin of chiral symmetry breaking. 
To identify the consequence of the chiral symmetry breaking to the Lefschetz thimbles must become a benchmark in the future study on applications of Lefschetz thimbles to the sign problem. 


\acknowledgments

Y.T. is financially supported by Special Postdoctoral Researchers program of RIKEN. M.T. is supported in part by the JSPS Grant-in-Aid for Scientific Research, Grant No.16K05357.

\bibliographystyle{utphys}
\bibliography{QFT,lefschetz,./ref}

\providecommand{\href}[2]{#2}\begingroup\raggedright\begin{thebibliography}{10}

\bibitem{PhysRevB.41.9301}
E.~Y. Loh, J.~E. Gubernatis, R.~T. Scalettar, S.~R. White, D.~J. Scalapino, and
  R.~L. Sugar, ``{Sign problem in the numerical simulation of many-electron
  systems},'' \href{http://dx.doi.org/10.1103/PhysRevB.41.9301}{{\em Phys. Rev.
  B} {\bfseries 41} (May, 1990) 9301--9307}.

\bibitem{Batrouni:1992fj}
G.~G. Batrouni and P.~de~Forcrand, ``{The Fermion sign problem: A New
  decoupling transformation, and a new simulation algorithm},''
  \href{http://dx.doi.org/10.1103/PhysRevB.48.589}{{\em Phys. Rev. B}
  {\bfseries 48} (1993) 589},
\href{http://arxiv.org/abs/cond-mat/9211009}{{\ttfamily arXiv:cond-mat/9211009
  [cond-mat]}}.

\bibitem{Muroya:2003qs}
S.~Muroya, A.~Nakamura, C.~Nonaka, and T.~Takaishi, ``{Lattice QCD at finite
  density: An Introductory review},''
  \href{http://dx.doi.org/10.1143/PTP.110.615}{{\em Prog. Theor. Phys.}
  {\bfseries 110} (2003) 615--668},
\href{http://arxiv.org/abs/hep-lat/0306031}{{\ttfamily arXiv:hep-lat/0306031
  [hep-lat]}}.

\bibitem{Schwinger:1962tn}
J.~S. Schwinger, ``{Gauge Invariance and Mass},''
\href{http://dx.doi.org/10.1103/PhysRev.125.397}{{\em Phys. Rev.} {\bfseries
  125} (1962) 397--398}.

\bibitem{Schwinger:1962tp}
J.~S. Schwinger, ``{Gauge Invariance and Mass. 2.},''
\href{http://dx.doi.org/10.1103/PhysRev.128.2425}{{\em Phys. Rev.} {\bfseries
  128} (1962) 2425--2429}.

\bibitem{Narayanan:2012du}
R.~Narayanan, ``{QED at a finite chemical potential},''
  \href{http://dx.doi.org/10.1103/PhysRevD.86.087701}{{\em Phys. Rev.}
  {\bfseries D86} (2012) 087701},
\href{http://arxiv.org/abs/1206.1489}{{\ttfamily arXiv:1206.1489 [hep-lat]}}.

\bibitem{Narayanan:2012qf}
R.~Narayanan, ``{Two flavor massless Schwinger model on a torus at a finite
  chemical potential},''
  \href{http://dx.doi.org/10.1103/PhysRevD.86.125008}{{\em Phys. Rev.}
  {\bfseries D86} (2012) 125008},
\href{http://arxiv.org/abs/1210.3072}{{\ttfamily arXiv:1210.3072 [hep-th]}}.

\bibitem{Lohmayer:2013eka}
R.~Lohmayer and R.~Narayanan, ``{Phase structure of two-dimensional QED at zero
  temperature with flavor-dependent chemical potentials and the role of
  multidimensional theta functions},''
  \href{http://dx.doi.org/10.1103/PhysRevD.88.105030}{{\em Phys. Rev.}
  {\bfseries D88} no.~10, (2013) 105030},
\href{http://arxiv.org/abs/1307.4969}{{\ttfamily arXiv:1307.4969 [hep-th]}}.

\bibitem{Gattringer:2015nea}
C.~Gattringer, T.~Kloiber, and V.~Sazonov, ``{Solving the sign problems of the
  massless lattice Schwinger model with a dual formulation},''
  \href{http://dx.doi.org/10.1016/j.nuclphysb.2015.06.017}{{\em Nucl. Phys.}
  {\bfseries B897} (2015) 732--748},
\href{http://arxiv.org/abs/1502.05479}{{\ttfamily arXiv:1502.05479 [hep-lat]}}.

\bibitem{Banuls:2016gid}
M.~C. Ba\~nuls, K.~Cichy, J.~I. Cirac, K.~Jansen, and S.~K\"uhn, ``{Density
  induced phase transitions in QED$_\mathrm{2}$ - A study with matrix product
  states},''
\href{http://arxiv.org/abs/1611.00705}{{\ttfamily arXiv:1611.00705 [hep-lat]}}.

\bibitem{pham1983vanishing}
F.~Pham, \href{http://dx.doi.org/10.1090/pspum/040.2}{``{Vanishing homologies
  and the $n$ variable saddlepoint method},''} in {\em Proc. Symp. Pure Math},
  vol.~40.2, pp.~319--333.
\newblock AMS, 1983.

\bibitem{Kaminski1994}
D.~Kaminski, ``{Exponentially improved stationary phase approximations for
  double integrals},'' \href{http://dx.doi.org/10.4310/MAA.1994.v1.n1.a4}{{\em
  Methods and Appl. of Analysis} {\bfseries 1} (1994) 44--56}.

\bibitem{Howls2271}
C.~J. Howls, ``Hyperasymptotics for multidimensional integrals, exact remainder
  terms and the global connection problem,''
  \href{http://dx.doi.org/10.1098/rspa.1997.0122}{{\em Proc. R. Soc. A}
  {\bfseries 453} no.~1966, (1997) 2271--2294}.

\bibitem{Witten:2010cx}
E.~Witten, ``{Analytic Continuation Of Chern-Simons Theory},'' in {\em
  {Chern-Simons Gauge Theory: 20 Years After}}, vol.~50, pp.~347--446.
\newblock AMS/IP Stud. Adv. Math., 2010.
\newblock
\href{http://arxiv.org/abs/1001.2933}{{\ttfamily arXiv:1001.2933 [hep-th]}}.
\newblock

\bibitem{Witten:2010zr}
E.~Witten, ``{A New Look At The Path Integral Of Quantum Mechanics},''
\href{http://arxiv.org/abs/1009.6032}{{\ttfamily arXiv:1009.6032 [hep-th]}}.

\bibitem{Harlow:2011ny}
D.~Harlow, J.~Maltz, and E.~Witten, ``{Analytic Continuation of Liouville
  Theory},'' \href{http://dx.doi.org/10.1007/JHEP12(2011)071}{{\em JHEP}
  {\bfseries 1112} (2011) 071},
\href{http://arxiv.org/abs/1108.4417}{{\ttfamily arXiv:1108.4417 [hep-th]}}.

\bibitem{Dunne:2012ae}
G.~V. Dunne and M.~\"Unsal, ``{Resurgence and Trans-series in Quantum Field
  Theory: The CP(N-1) Model},''
  \href{http://dx.doi.org/10.1007/JHEP11(2012)170}{{\em JHEP} {\bfseries 1211}
  (2012) 170},
\href{http://arxiv.org/abs/1210.2423}{{\ttfamily arXiv:1210.2423 [hep-th]}}.

\bibitem{Basar:2013eka}
G.~Basar, G.~V. Dunne, and M.~\"Unsal, ``{Resurgence theory, ghost-instantons,
  and analytic continuation of path integrals},''
  \href{http://dx.doi.org/10.1007/JHEP10(2013)041}{{\em JHEP} {\bfseries 1310}
  (2013) 041},
\href{http://arxiv.org/abs/1308.1108}{{\ttfamily arXiv:1308.1108 [hep-th]}}.

\bibitem{Cherman:2014ofa}
A.~Cherman, D.~Dorigoni, and M.~\"Unsal, ``{Decoding perturbation theory using
  resurgence: Stokes phenomena, new saddle points and Lefschetz thimbles},''
  \href{http://dx.doi.org/10.1007/JHEP10(2015)056}{{\em JHEP} {\bfseries 10}
  (2015) 056},
\href{http://arxiv.org/abs/1403.1277}{{\ttfamily arXiv:1403.1277 [hep-th]}}.

\bibitem{Cherman:2014xia}
A.~Cherman, P.~Koroteev, and M.~\"Unsal, ``{Resurgence and Holomorphy: From
  Weak to Strong Coupling},'' \href{http://dx.doi.org/10.1063/1.4921155}{{\em
  J. Math. Phys.} {\bfseries 56} no.~5, (2015) 053505},
\href{http://arxiv.org/abs/1410.0388}{{\ttfamily arXiv:1410.0388 [hep-th]}}.

\bibitem{Dorigoni:2014hea}
D.~Dorigoni, ``{An Introduction to Resurgence, Trans-Series and Alien
  Calculus},''
\href{http://arxiv.org/abs/1411.3585}{{\ttfamily arXiv:1411.3585 [hep-th]}}.

\bibitem{David:1992za}
F.~David, ``{Nonperturbative effects in matrix models and vacua of
  two-dimensional gravity},''
  \href{http://dx.doi.org/10.1016/0370-2693(93)90417-G}{{\em Phys.Lett.}
  {\bfseries B302} (1993) 403--410},
\href{http://arxiv.org/abs/hep-th/9212106}{{\ttfamily arXiv:hep-th/9212106
  [hep-th]}}.

\bibitem{Felder:2004uy}
G.~Felder and R.~Riser, ``{Holomorphic matrix integrals},''
  \href{http://dx.doi.org/10.1016/j.nuclphysb.2004.05.010}{{\em Nucl.Phys.}
  {\bfseries B691} (2004) 251--258},
\href{http://arxiv.org/abs/hep-th/0401191}{{\ttfamily arXiv:hep-th/0401191
  [hep-th]}}.

\bibitem{Marino:2008ya}
M.~Marino, ``{Nonperturbative effects and nonperturbative definitions in matrix
  models and topological strings},''
  \href{http://dx.doi.org/10.1088/1126-6708/2008/12/114}{{\em JHEP} {\bfseries
  12} (2008) 114},
\href{http://arxiv.org/abs/0805.3033}{{\ttfamily arXiv:0805.3033 [hep-th]}}.

\bibitem{Marino:2012zq}
M.~Mari\~{n}o, ``{Lectures on non-perturbative effects in large $N$ gauge
  theories, matrix models and strings},''
  \href{http://dx.doi.org/10.1002/prop.201400005}{{\em Fortsch. Phys.}
  {\bfseries 62} (2014) 455--540},
\href{http://arxiv.org/abs/1206.6272}{{\ttfamily arXiv:1206.6272 [hep-th]}}.

\bibitem{Schiappa:2013opa}
R.~Schiappa and R.~Vaz, ``{The Resurgence of Instantons: Multi-Cut Stokes
  Phases and the Painleve II Equation},''
  \href{http://dx.doi.org/10.1007/s00220-014-2028-7}{{\em Commun. Math. Phys.}
  {\bfseries 330} (2014) 655--721},
\href{http://arxiv.org/abs/1302.5138}{{\ttfamily arXiv:1302.5138 [hep-th]}}.

\bibitem{Behtash:2015kna}
A.~Behtash, T.~Sulejmanpasic, T.~Sch{\"a}fer, and M.~{\"U}nsal, ``{Hidden
  Topological Angles in Path Integrals},''
  \href{http://dx.doi.org/10.1103/PhysRevLett.115.041601}{{\em Phys. Rev.
  Lett.} {\bfseries 115} no.~4, (2015) 041601},
\href{http://arxiv.org/abs/1502.06624}{{\ttfamily arXiv:1502.06624 [hep-th]}}.

\bibitem{Behtash:2015kva}
A.~Behtash, E.~Poppitz, T.~Sulejmanpasic, and M.~{\"U}nsal, ``{The curious
  incident of multi-instantons and the necessity of Lefschetz thimbles},''
  \href{http://dx.doi.org/10.1007/JHEP11(2015)175}{{\em JHEP} {\bfseries 11}
  (2015) 175},
\href{http://arxiv.org/abs/1507.04063}{{\ttfamily arXiv:1507.04063 [hep-th]}}.

\bibitem{Gukov:2016njj}
S.~Gukov, M.~Marino, and P.~Putrov, ``{Resurgence in complex Chern-Simons
  theory},''
\href{http://arxiv.org/abs/1605.07615}{{\ttfamily arXiv:1605.07615 [hep-th]}}.

\bibitem{Gukov:2016tnp}
S.~Gukov, ``{RG Flows and Bifurcations},''
\href{http://arxiv.org/abs/1608.06638}{{\ttfamily arXiv:1608.06638 [hep-th]}}.

\bibitem{Fujimori:2016ljw}
T.~Fujimori, S.~Kamata, T.~Misumi, M.~Nitta, and N.~Sakai, ``{Nonperturbative
  contributions from complexified solutions in $\mathbb{C}P^{N-1}$models},''
  \href{http://dx.doi.org/10.1103/PhysRevD.94.105002}{{\em Phys. Rev.}
  {\bfseries D94} no.~10, (2016) 105002},
\href{http://arxiv.org/abs/1607.04205}{{\ttfamily arXiv:1607.04205 [hep-th]}}.

\bibitem{Kozcaz:2016wvy}
C.~Koz\c{c}az, T.~Sulejmanpasic, Y.~Tanizaki, and M.~\"Unsal, ``{Cheshire Cat
  resurgence, Self-resurgence and Quasi-Exact Solvable Systems},''
\href{http://arxiv.org/abs/1609.06198}{{\ttfamily arXiv:1609.06198 [hep-th]}}.

\bibitem{Cristoforetti:2012su}
{\bfseries AuroraScience} Collaboration, M.~Cristoforetti, F.~Di~Renzo, and
  L.~Scorzato, ``{New approach to the sign problem in quantum field theories:
  High density QCD on a Lefschetz thimble},''
  \href{http://dx.doi.org/10.1103/PhysRevD.86.074506}{{\em Phys. Rev. D}
  {\bfseries 86} (2012) 074506},
\href{http://arxiv.org/abs/1205.3996}{{\ttfamily arXiv:1205.3996 [hep-lat]}}.

\bibitem{Cristoforetti:2013wha}
M.~Cristoforetti, F.~Di~Renzo, A.~Mukherjee, and L.~Scorzato, ``{Monte Carlo
  simulations on the Lefschetz thimble: taming the sign problem},''
  \href{http://dx.doi.org/10.1103/PhysRevD.88.051501}{{\em Phys. Rev. D}
  {\bfseries 88} (2013) 051501},
\href{http://arxiv.org/abs/1303.7204}{{\ttfamily arXiv:1303.7204 [hep-lat]}}.

\bibitem{Cristoforetti:2014gsa}
M.~Cristoforetti, F.~Di~Renzo, G.~Eruzzi, A.~Mukherjee, C.~Schmidt,
  L.~Scorzato, and C.~Torrero, ``{An efficient method to compute the residual
  phase on a Lefschetz thimble},''
  \href{http://dx.doi.org/10.1103/PhysRevD.89.114505}{{\em Phys. Rev. D}
  {\bfseries 89} (2014) 114505},
\href{http://arxiv.org/abs/1403.5637}{{\ttfamily arXiv:1403.5637 [hep-lat]}}.

\bibitem{Aarts:2013fpa}
G.~Aarts, ``{Lefschetz thimbles and stochastic quantisation: Complex actions in
  the complex plane},''
  \href{http://dx.doi.org/10.1103/PhysRevD.88.094501}{{\em Phys. Rev. D}
  {\bfseries 88} (2013) 094501},
\href{http://arxiv.org/abs/1308.4811}{{\ttfamily arXiv:1308.4811 [hep-lat]}}.

\bibitem{Fujii:2013sra}
H.~Fujii, D.~Honda, M.~Kato, Y.~Kikukawa, S.~Komatsu, and T.~Sano, ``{Hybrid
  Monte Carlo on Lefschetz thimbles - A study of the residual sign problem},''
  \href{http://dx.doi.org/10.1007/JHEP10(2013)147}{{\em JHEP} {\bfseries 1310}
  (2013) 147},
\href{http://arxiv.org/abs/1309.4371}{{\ttfamily arXiv:1309.4371 [hep-lat]}}.

\bibitem{Mukherjee:2014hsa}
A.~Mukherjee and M.~Cristoforetti, ``{Lefschetz thimble Monte Carlo for many
  body theories: application to the repulsive Hubbard model away from half
  filling},'' \href{http://dx.doi.org/10.1103/PhysRevB.90.035134}{{\em Phys.
  Rev. B} {\bfseries 90} (2014) 035134},
\href{http://arxiv.org/abs/1403.5680}{{\ttfamily arXiv:1403.5680
  [cond-mat.str-el]}}.

\bibitem{Aarts:2014nxa}
G.~Aarts, L.~Bongiovanni, E.~Seiler, and D.~Sexty, ``{Some remarks on Lefschetz
  thimbles and complex Langevin dynamics},''
  \href{http://dx.doi.org/10.1007/JHEP10(2014)159}{{\em JHEP} {\bfseries 1410}
  (2014) 159},
\href{http://arxiv.org/abs/1407.2090}{{\ttfamily arXiv:1407.2090 [hep-lat]}}.

\bibitem{Tanizaki:2014xba}
Y.~Tanizaki and T.~Koike, ``{Real-time Feynman path integral with
  Picard--Lefschetz theory and its applications to quantum tunneling},''
  \href{http://dx.doi.org/10.1016/j.aop.2014.09.003}{{\em Ann. Phys.}
  {\bfseries 351} (2014) 250},
\href{http://arxiv.org/abs/1406.2386}{{\ttfamily arXiv:1406.2386 [math-ph]}}.

\bibitem{Cherman:2014sba}
A.~Cherman and M.~Unsal, ``{Real-Time Feynman Path Integral Realization of
  Instantons},''
\href{http://arxiv.org/abs/1408.0012}{{\ttfamily arXiv:1408.0012 [hep-th]}}.

\bibitem{Tanizaki:2014tua}
Y.~Tanizaki, ``{Lefschetz-thimble techniques for path integral of
  zero-dimensional $O(n)$ sigma models},''
  \href{http://dx.doi.org/10.1103/PhysRevD.91.036002}{{\em Phys. Rev. D}
  {\bfseries 91} (2015) 036002},
\href{http://arxiv.org/abs/1412.1891}{{\ttfamily arXiv:1412.1891 [hep-th]}}.

\bibitem{Kanazawa:2014qma}
T.~Kanazawa and Y.~Tanizaki, ``{Structure of Lefschetz thimbles in simple
  fermionic systems},'' \href{http://dx.doi.org/10.1007/JHEP03(2015)044}{{\em
  JHEP} {\bfseries 1503} (2015) 044},
\href{http://arxiv.org/abs/1412.2802}{{\ttfamily arXiv:1412.2802 [hep-th]}}.

\bibitem{Tanizaki:2015pua}
Y.~Tanizaki, H.~Nishimura, and K.~Kashiwa, ``{Evading the sign problem in the
  mean-field approximation through Lefschetz-thimble path integral},''
  \href{http://dx.doi.org/10.1103/PhysRevD.91.101701}{{\em Phys. Rev. D}
  {\bfseries 91} (2015) 101701},
\href{http://arxiv.org/abs/1504.02979}{{\ttfamily arXiv:1504.02979 [hep-th]}}.

\bibitem{DiRenzo:2015foa}
F.~Di~Renzo and G.~Eruzzi, ``{Thimble regularization at work: from toy models
  to chiral random matrix theories},''
  \href{http://dx.doi.org/10.1103/PhysRevD.92.085030}{{\em Phys. Rev. D}
  {\bfseries 92} (2015) 085030},
\href{http://arxiv.org/abs/1507.03858}{{\ttfamily arXiv:1507.03858 [hep-lat]}}.

\bibitem{Fukushima:2015qza}
K.~Fukushima and Y.~Tanizaki, ``{Hamilton dynamics for the Lefschetz thimble
  integration akin to the complex Langevin method},''
  \href{http://dx.doi.org/10.1093/ptep/ptv152}{{\em Prog. Theor. Exp. Phys.}
  {\bfseries 2015} (2015) 111A01},
\href{http://arxiv.org/abs/1507.07351}{{\ttfamily arXiv:1507.07351 [hep-th]}}.

\bibitem{Tsutsui:2015tua}
S.~Tsutsui and T.~M. Doi, ``{An improvement in complex Langevin dynamics from a
  view point of Lefschetz thimbles},''
  \href{http://dx.doi.org/10.1103/PhysRevD.94.074009}{{\em Phys. Rev.}
  {\bfseries D94} (2016) 074009},
\href{http://arxiv.org/abs/1508.04231}{{\ttfamily arXiv:1508.04231 [hep-lat]}}.

\bibitem{Tanizaki:2015rda}
Y.~Tanizaki, Y.~Hidaka, and T.~Hayata, ``{Lefschetz-thimble analysis of the
  sign problem in one-site fermion model},''
  \href{http://dx.doi.org/10.1088/1367-2630/18/3/033002}{{\em New J. Phys.}
  {\bfseries 18} (2016) 033002},
\href{http://arxiv.org/abs/1509.07146}{{\ttfamily arXiv:1509.07146 [hep-th]}}.

\bibitem{Fujii:2015bua}
H.~Fujii, S.~Kamata, and Y.~Kikukawa, ``{Lefschetz thimble structure in
  one-dimensional lattice Thirring model at finite density},''
  \href{http://dx.doi.org/10.1007/JHEP02(2016)036,
  10.1007/JHEP11(2015)078}{{\em JHEP} {\bfseries 11} (2015) 078},
  \href{http://arxiv.org/abs/1509.08176}{{\ttfamily arXiv:1509.08176
  [hep-lat]}}.
[Erratum: JHEP02,036(2016)].

\bibitem{Fujii:2015vha}
H.~Fujii, S.~Kamata, and Y.~Kikukawa, ``{Monte Carlo study of Lefschetz thimble
  structure in one-dimensional Thirring model at finite density},''
  \href{http://dx.doi.org/10.1007/JHEP12(2015)125}{{\em JHEP} {\bfseries 12}
  (2015) 125},
\href{http://arxiv.org/abs/1509.09141}{{\ttfamily arXiv:1509.09141 [hep-lat]}}.

\bibitem{Alexandru:2015xva}
A.~Alexandru, G.~Basar, and P.~Bedaque, ``{Monte Carlo algorithm for simulating
  fermions on Lefschetz thimbles},''
  \href{http://dx.doi.org/10.1103/PhysRevD.93.014504}{{\em Phys. Rev.}
  {\bfseries D93} (2016) 014504},
\href{http://arxiv.org/abs/1510.03258}{{\ttfamily arXiv:1510.03258 [hep-lat]}}.

\bibitem{Hayata:2015lzj}
T.~Hayata, Y.~Hidaka, and Y.~Tanizaki, ``{Complex saddle points and the sign
  problem in complex Langevin simulation},''
  \href{http://dx.doi.org/10.1016/j.nuclphysb.2016.07.031}{{\em Nucl. Phys.}
  {\bfseries B911} (2016) 94--105},
\href{http://arxiv.org/abs/1511.02437}{{\ttfamily arXiv:1511.02437 [hep-lat]}}.

\bibitem{Alexandru:2015sua}
A.~Alexandru, G.~Basar, P.~F. Bedaque, G.~W. Ridgway, and N.~C. Warrington,
  ``{Sign problem and Monte Carlo calculations beyond Lefschetz thimbles},''
  \href{http://dx.doi.org/10.1007/JHEP05(2016)053}{{\em JHEP} {\bfseries 05}
  (2016) 053},
\href{http://arxiv.org/abs/1512.08764}{{\ttfamily arXiv:1512.08764 [hep-lat]}}.

\bibitem{Alexandru:2016gsd}
A.~Alexandru, G.~Basar, P.~F. Bedaque, S.~Vartak, and N.~C. Warrington,
  ``{Monte Carlo Study of Real Time Dynamics on the Lattice},''
  \href{http://dx.doi.org/10.1103/PhysRevLett.117.081602}{{\em Phys. Rev.
  Lett.} {\bfseries 117} (2016) 081602},
\href{http://arxiv.org/abs/1605.08040}{{\ttfamily arXiv:1605.08040 [hep-lat]}}.

\bibitem{Alexandru:2016ejd}
A.~Alexandru, G.~Basar, P.~F. Bedaque, G.~W. Ridgway, and N.~C. Warrington,
  ``{Monte Carlo calculations of the finite density Thirring model},''
  \href{http://dx.doi.org/10.1103/PhysRevD.95.014502}{{\em Phys. Rev.}
  {\bfseries D95} (2017) 014502},
\href{http://arxiv.org/abs/1609.01730}{{\ttfamily arXiv:1609.01730 [hep-lat]}}.

\bibitem{Dumlu:2010ua}
C.~K. Dumlu and G.~V. Dunne, ``{The Stokes Phenomenon and Schwinger Vacuum Pair
  Production in Time-Dependent Laser Pulses},''
  \href{http://dx.doi.org/10.1103/PhysRevLett.104.250402}{{\em Phys. Rev.
  Lett.} {\bfseries 104} (2010) 250402},
\href{http://arxiv.org/abs/1004.2509}{{\ttfamily arXiv:1004.2509 [hep-th]}}.

\bibitem{Dumlu:2011rr}
C.~K. Dumlu and G.~V. Dunne, ``{Interference Effects in Schwinger Vacuum Pair
  Production for Time-Dependent Laser Pulses},''
  \href{http://dx.doi.org/10.1103/PhysRevD.83.065028}{{\em Phys. Rev. D}
  {\bfseries 83} (2011) 065028},
\href{http://arxiv.org/abs/1102.2899}{{\ttfamily arXiv:1102.2899 [hep-th]}}.

\bibitem{Dumlu:2011cc}
C.~K. Dumlu and G.~V. Dunne, ``{Complex Worldline Instantons and Quantum
  Interference in Vacuum Pair Production},''
  \href{http://dx.doi.org/10.1103/PhysRevD.84.125023}{{\em Phys. Rev. D}
  {\bfseries 84} (2011) 125023},
\href{http://arxiv.org/abs/1110.1657}{{\ttfamily arXiv:1110.1657 [hep-th]}}.

\bibitem{Buividovich:2015oju}
P.~V. Buividovich, G.~V. Dunne, and S.~N. Valgushev, ``{Complex Path Integrals
  and Saddles in Two-Dimensional Gauge Theory},''
  \href{http://dx.doi.org/10.1103/PhysRevLett.116.132001}{{\em Phys. Rev.
  Lett.} {\bfseries 116} no.~13, (2016) 132001},
\href{http://arxiv.org/abs/1512.09021}{{\ttfamily arXiv:1512.09021 [hep-th]}}.

\bibitem{Alvarez:2016rmo}
G.~\'Alvarez, L.~Mart\'inez~Alonso, and E.~Medina, ``{Complex saddles in the
  Gross-Witten-Wadia matrix model},''
  \href{http://dx.doi.org/10.1103/PhysRevD.94.105010}{{\em Phys. Rev.}
  {\bfseries D94} no.~10, (2016) 105010},
\href{http://arxiv.org/abs/1610.09948}{{\ttfamily arXiv:1610.09948 [hep-th]}}.

\bibitem{Sachs:1991en}
I.~Sachs and A.~Wipf, ``{Finite temperature Schwinger model},'' {\em Helv.
  Phys. Acta} {\bfseries 65} (1992) 652--678,
\href{http://arxiv.org/abs/1005.1822}{{\ttfamily arXiv:1005.1822 [hep-th]}}.

\bibitem{Langfeld:2011rh}
K.~Langfeld and A.~Wipf, ``{Fermi-Einstein condensation in dense QCD-like
  theories},'' \href{http://dx.doi.org/10.1016/j.aop.2011.11.020}{{\em Annals
  Phys.} {\bfseries 327} (2012) 994--1029},
\href{http://arxiv.org/abs/1109.0502}{{\ttfamily arXiv:1109.0502 [hep-lat]}}.

\bibitem{KorthalsAltes:1993ca}
C.~Korthals~Altes, ``{Constrained effective potential in hot QCD},''
  \href{http://dx.doi.org/10.1016/0550-3213(94)90081-7}{{\em Nucl.Phys.}
  {\bfseries B420} (1994) 637--668},
\href{http://arxiv.org/abs/hep-th/9310195}{{\ttfamily arXiv:hep-th/9310195
  [hep-th]}}.

\bibitem{Fukuda:1974ey}
R.~Fukuda and E.~Kyriakopoulos, ``{Derivation of the Effective Potential},''
\href{http://dx.doi.org/10.1016/0550-3213(75)90014-0}{{\em Nucl.Phys.}
  {\bfseries B85} (1975) 354}.

\bibitem{Dumitru:2005ng}
A.~Dumitru, R.~D. Pisarski, and D.~Zschiesche, ``{Dense quarks, and the fermion
  sign problem, in a SU(N) matrix model},''
  \href{http://dx.doi.org/10.1103/PhysRevD.72.065008}{{\em Phys. Rev. D}
  {\bfseries 72} (2005) 065008},
\href{http://arxiv.org/abs/hep-ph/0505256}{{\ttfamily arXiv:hep-ph/0505256
  [hep-ph]}}.

\bibitem{Fukushima:2006uv}
K.~Fukushima and Y.~Hidaka, ``{A Model study of the sign problem in the
  mean-field approximation},''
  \href{http://dx.doi.org/10.1103/PhysRevD.75.036002}{{\em Phys. Rev. D}
  {\bfseries 75} (2007) 036002},
\href{http://arxiv.org/abs/hep-ph/0610323}{{\ttfamily arXiv:hep-ph/0610323
  [hep-ph]}}.

\bibitem{Nishimura:2014rxa}
H.~Nishimura, M.~C. Ogilvie, and K.~Pangeni, ``{Complex saddle points in QCD at
  finite temperature and density},''
  \href{http://dx.doi.org/10.1103/PhysRevD.90.045039}{{\em Phys. Rev. D}
  {\bfseries 90} (2014) 045039},
\href{http://arxiv.org/abs/1401.7982}{{\ttfamily arXiv:1401.7982 [hep-ph]}}.

\bibitem{Nishimura:2014kla}
H.~Nishimura, M.~C. Ogilvie, and K.~Pangeni, ``{Complex Saddle Points and
  Disorder Lines in QCD at finite temperature and density},''
  \href{http://dx.doi.org/10.1103/PhysRevD.91.054004}{{\em Phys. Rev. D}
  {\bfseries 91} no.~5, (2015) 054004},
\href{http://arxiv.org/abs/1411.4959}{{\ttfamily arXiv:1411.4959 [hep-ph]}}.

\bibitem{Fukushima:2003fw}
K.~Fukushima, ``{Chiral effective model with the Polyakov loop},''
  \href{http://dx.doi.org/10.1016/j.physletb.2004.04.027}{{\em Phys.Lett.}
  {\bfseries B591} (2004) 277--284},
\href{http://arxiv.org/abs/hep-ph/0310121}{{\ttfamily arXiv:hep-ph/0310121
  [hep-ph]}}.

\bibitem{Alexandrou:1998wv}
C.~Alexandrou, A.~Borici, A.~Feo, P.~de~Forcrand, A.~Galli, F.~Jegerlehner, and
  T.~Takaishi, ``{The Deconfinement phase transition in one flavor QCD},''
  \href{http://dx.doi.org/10.1103/PhysRevD.60.034504}{{\em Phys.Rev.}
  {\bfseries D60} (1999) 034504},
\href{http://arxiv.org/abs/hep-lat/9811028}{{\ttfamily arXiv:hep-lat/9811028
  [hep-lat]}}.

\bibitem{Condella:1999bk}
J.~Condella and C.~E. Detar, ``{Potts flux tube model at nonzero chemical
  potential},'' \href{http://dx.doi.org/10.1103/PhysRevD.61.074023}{{\em
  Phys.Rev.} {\bfseries D61} (2000) 074023},
\href{http://arxiv.org/abs/hep-lat/9910028}{{\ttfamily arXiv:hep-lat/9910028
  [hep-lat]}}.

\bibitem{Alford:2001ug}
M.~G. Alford, S.~Chandrasekharan, J.~Cox, and U.~Wiese, ``{Solution of the
  complex action problem in the Potts model for dense QCD},''
  \href{http://dx.doi.org/10.1016/S0550-3213(01)00068-2}{{\em Nucl.Phys.}
  {\bfseries B602} (2001) 61--86},
\href{http://arxiv.org/abs/hep-lat/0101012}{{\ttfamily arXiv:hep-lat/0101012
  [hep-lat]}}.

\bibitem{Banks:1983me}
T.~Banks and A.~Ukawa, ``{Deconfining and Chiral Phase Transitions in Quantum
  Chromodynamics at Finite Temperature},''
\href{http://dx.doi.org/10.1016/0550-3213(83)90016-0}{{\em Nucl.Phys.}
  {\bfseries B225} (1983) 145}.

\bibitem{Pisarski:2000eq}
R.~D. Pisarski, ``{Quark gluon plasma as a condensate of SU(3) Wilson lines},''
  \href{http://dx.doi.org/10.1103/PhysRevD.62.111501}{{\em Phys.Rev.}
  {\bfseries D62} (2000) 111501},
\href{http://arxiv.org/abs/hep-ph/0006205}{{\ttfamily arXiv:hep-ph/0006205
  [hep-ph]}}.

\bibitem{Dumitru:2000in}
A.~Dumitru and R.~D. Pisarski, ``{Event-by-event fluctuations from decay of a
  Polyakov loop condensate},''
  \href{http://dx.doi.org/10.1016/S0370-2693(01)00286-6}{{\em Phys.Lett.}
  {\bfseries B504} (2001) 282--290},
\href{http://arxiv.org/abs/hep-ph/0010083}{{\ttfamily arXiv:hep-ph/0010083
  [hep-ph]}}.

\bibitem{Akerlund:2016myr}
O.~Akerlund, P.~de~Forcrand, and T.~Rindlisbacher, ``{Oscillating propagators
  in heavy-dense QCD},'' \href{http://dx.doi.org/10.1007/JHEP10(2016)055}{{\em
  JHEP} {\bfseries 10} (2016) 055},
\href{http://arxiv.org/abs/1602.02925}{{\ttfamily arXiv:1602.02925 [hep-lat]}}.

\bibitem{Hirakida:2016rqd}
T.~Hirakida, H.~Kouno, J.~Takahashi, and M.~Yahiro, ``{Interplay between sign
  problem and $Z_3$ symmetry in three-dimensional Potts models},''
  \href{http://dx.doi.org/10.1103/PhysRevD.94.014011}{{\em Phys. Rev.}
  {\bfseries D94} no.~1, (2016) 014011},
\href{http://arxiv.org/abs/1604.02977}{{\ttfamily arXiv:1604.02977 [hep-lat]}}.

\bibitem{Bender:1992gn}
I.~Bender, T.~Hashimoto, F.~Karsch, V.~Linke, A.~Nakamura, M.~Plewnia, I.~O.
  Stamatescu, and W.~Wetzel, ``{Full QCD and QED at finite temperature and
  chemical potential},''
\href{http://dx.doi.org/10.1016/0920-5632(92)90265-T}{{\em Nucl. Phys. Proc.
  Suppl.} {\bfseries 26} (1992) 323--325}.

\bibitem{Blum:1995cb}
T.~C. Blum, J.~E. Hetrick, and D.~Toussaint, ``{High density QCD with static
  quarks},'' \href{http://dx.doi.org/10.1103/PhysRevLett.76.1019}{{\em Phys.
  Rev. Lett.} {\bfseries 76} (1996) 1019--1022},
\href{http://arxiv.org/abs/hep-lat/9509002}{{\ttfamily arXiv:hep-lat/9509002
  [hep-lat]}}.

\bibitem{Hands:2010zp}
S.~Hands, T.~J. Hollowood, and J.~C. Myers, ``{QCD with Chemical Potential in a
  Small Hyperspherical Box},''
  \href{http://dx.doi.org/10.1007/JHEP07(2010)086}{{\em JHEP} {\bfseries 07}
  (2010) 086},
\href{http://arxiv.org/abs/1003.5813}{{\ttfamily arXiv:1003.5813 [hep-th]}}.

\bibitem{Reinosa:2015oua}
U.~Reinosa, J.~Serreau, and M.~Tissier, ``{Perturbative study of the QCD phase
  diagram for heavy quarks at nonzero chemical potential},''
  \href{http://dx.doi.org/10.1103/PhysRevD.92.025021}{{\em Phys. Rev. D}
  {\bfseries 92} no.~2, (2015) 025021},
\href{http://arxiv.org/abs/1504.02916}{{\ttfamily arXiv:1504.02916 [hep-th]}}.

\bibitem{Pawlowski:2013pje}
J.~M. Pawlowski and C.~Zielinski, ``{Thirring model at finite density in 0+1
  dimensions with stochastic quantization: Crosscheck with an exact
  solution},'' \href{http://dx.doi.org/10.1103/PhysRevD.87.094503}{{\em Phys.
  Rev. D} {\bfseries 87} (2013) 094503},
\href{http://arxiv.org/abs/1302.1622}{{\ttfamily arXiv:1302.1622 [hep-lat]}}.

\bibitem{Coleman:1973ci}
S.~R. Coleman, ``{There are no Goldstone bosons in two-dimensions},''
\href{http://dx.doi.org/10.1007/BF01646487}{{\em Commun. Math. Phys.}
  {\bfseries 31} (1973) 259--264}.

\bibitem{mermin1966absence}
N.~D. Mermin and H.~Wagner, ``{Absence of ferromagnetism or antiferromagnetism
  in one-or two-dimensional isotropic Heisenberg models},''
  \href{http://dx.doi.org/10.1103/PhysRevLett.17.1133}{{\em Phys.~Rev.~Lett.}
  {\bfseries 17} (1966) 1133}.

\bibitem{Gibbs:1986ut}
P.~E. Gibbs, ``{Lattice Monte Carlo Simulations of {QCD} at Finite Baryonic
  Density},''
\href{http://dx.doi.org/10.1016/0370-2693(86)90109-7}{{\em Phys. Lett.}
  {\bfseries B182} (1986) 369--372}.

\bibitem{Gibbs:1986hi}
P.~E. Gibbs, ``{The Fermion Propagator Matrix in Lattice {QCD}},''
\href{http://dx.doi.org/10.1016/0370-2693(86)90215-7}{{\em Phys. Lett.}
  {\bfseries B172} (1986) 53--61}.

\bibitem{Barbour:1997bh}
I.~M. Barbour, S.~E. Morrison, E.~G. Klepfish, J.~B. Kogut, and M.-P. Lombardo,
  ``{The Critical points of strongly coupled lattice QCD at nonzero chemical
  potential},'' \href{http://dx.doi.org/10.1103/PhysRevD.56.7063}{{\em Phys.
  Rev.} {\bfseries D56} (1997) 7063--7072},
\href{http://arxiv.org/abs/hep-lat/9705038}{{\ttfamily arXiv:hep-lat/9705038
  [hep-lat]}}.

\bibitem{Barbour:1997ej}
I.~M. Barbour, S.~E. Morrison, E.~G. Klepfish, J.~B. Kogut, and M.-P. Lombardo,
  ``{Results on finite density QCD},''
  \href{http://dx.doi.org/10.1016/S0920-5632(97)00484-2}{{\em Nucl. Phys. Proc.
  Suppl.} {\bfseries 60A} (1998) 220--234},
\href{http://arxiv.org/abs/hep-lat/9705042}{{\ttfamily arXiv:hep-lat/9705042
  [hep-lat]}}.

\bibitem{Stephanov:1996ki}
M.~A. Stephanov, ``{Random matrix model of QCD at finite density and the nature
  of the quenched limit},''
  \href{http://dx.doi.org/10.1103/PhysRevLett.76.4472}{{\em Phys. Rev. Lett.}
  {\bfseries 76} (1996) 4472--4475},
\href{http://arxiv.org/abs/hep-lat/9604003}{{\ttfamily arXiv:hep-lat/9604003
  [hep-lat]}}.

\bibitem{Cohen:2003kd}
T.~D. Cohen, ``{Functional integrals for QCD at nonzero chemical potential and
  zero density},'' \href{http://dx.doi.org/10.1103/PhysRevLett.91.222001}{{\em
  Phys. Rev. Lett.} {\bfseries 91} (2003) 222001},
\href{http://arxiv.org/abs/hep-ph/0307089}{{\ttfamily arXiv:hep-ph/0307089
  [hep-ph]}}.

\bibitem{Splittorff:2006fu}
K.~Splittorff and J.~J.~M. Verbaarschot, ``{Phase of the Fermion Determinant at
  Nonzero Chemical Potential},''
  \href{http://dx.doi.org/10.1103/PhysRevLett.98.031601}{{\em Phys. Rev. Lett.}
  {\bfseries 98} (2007) 031601},
\href{http://arxiv.org/abs/hep-lat/0609076}{{\ttfamily arXiv:hep-lat/0609076
  [hep-lat]}}.

\bibitem{Splittorff:2007ck}
K.~Splittorff and J.~J.~M. Verbaarschot, ``{The QCD Sign Problem for Small
  Chemical Potential},''
  \href{http://dx.doi.org/10.1103/PhysRevD.75.116003}{{\em Phys. Rev.}
  {\bfseries D75} (2007) 116003},
\href{http://arxiv.org/abs/hep-lat/0702011}{{\ttfamily arXiv:hep-lat/0702011
  [HEP-LAT]}}.

\bibitem{Cherman:2010jj}
A.~Cherman, M.~Hanada, and D.~Robles-Llana, ``{Orbifold equivalence and the
  sign problem at finite baryon density},''
  \href{http://dx.doi.org/10.1103/PhysRevLett.106.091603}{{\em Phys. Rev.
  Lett.} {\bfseries 106} (2011) 091603},
\href{http://arxiv.org/abs/1009.1623}{{\ttfamily arXiv:1009.1623 [hep-th]}}.

\bibitem{Hidaka:2011jj}
Y.~Hidaka and N.~Yamamoto, ``{No-Go Theorem for Critical Phenomena in Large-Nc
  QCD},'' \href{http://dx.doi.org/10.1103/PhysRevLett.108.121601}{{\em Phys.
  Rev. Lett.} {\bfseries 108} (2012) 121601},
\href{http://arxiv.org/abs/1110.3044}{{\ttfamily arXiv:1110.3044 [hep-ph]}}.

\bibitem{Nagata:2012tc}
{\bfseries XQCD-J} Collaboration, K.~Nagata, S.~Motoki, Y.~Nakagawa,
  A.~Nakamura, and T.~Saito, ``{Towards extremely dense matter on the
  lattice},'' \href{http://dx.doi.org/10.1093/ptep/pts003}{{\em Prog. Theor.
  Exp. Phys.} {\bfseries 2012} (2012) 01A103},
\href{http://arxiv.org/abs/1204.1412}{{\ttfamily arXiv:1204.1412 [hep-lat]}}.

\end{thebibliography}\endgroup
\end{document}